\documentclass[a4paper,fleqn]{cas-sc}
\usepackage[authoryear]{natbib}
\usepackage{graphicx} 
\usepackage{float}
\usepackage{algorithm}  
\usepackage{algpseudocode}
\usepackage{color}
\usepackage{setspace}
\usepackage{cite}
\usepackage{amsmath,amssymb,amsfonts}
\usepackage{textcomp}
\usepackage{xcolor}
\usepackage{booktabs}
\usepackage{multirow}
\usepackage[switch]{lineno}
\usepackage{bbding}
\usepackage{pifont}
\usepackage{url}
\usepackage[nomarkers,figuresonly]{endfloat}

\usepackage{lineno}

\begin{document}
\let\WriteBookmarks\relax
\def\floatpagepagefraction{1}
\def\textpagefraction{.001}
\shorttitle{EFKAN}
\shortauthors{F.~Wang, H.~Qiu, Y.~Huang, X.~Gu, R.~Wang, and B.~Yang}

\title [mode = title]{EFKAN: A KAN-Integrated Neural Operator For Efficient Magnetotelluric Forward Modeling}

\author[1]{Feng~Wang}

\author[1]{Hong~Qiu}

\author[1]{Yingying~Huang}

\author[1]{Xiaozhe~Gu}

\author[1]{Renfang~Wang}

\author[2]{Bo~Yang}

\address[1]{College of Big Data and Software Engineering, Zhejiang Wanli University, Ningbo, 315200, China}
\address[2]{Key Laboratory of Geoscience Big Data and Deep Resource of Zhejiang Province, School of Earth Sciences, Zhejiang University, Hangzhou, 310027, China}

\begin{abstract}
Forward modeling is the cornerstone of magnetotelluric (MT) inversion. Neural operators have been successfully applied to solve partial differential equations, demonstrating encouraging performance in rapid MT forward modeling. In particular, they can obtain the electromagnetic field at arbitrary locations and frequencies, which is meaningful for MT forward modeling. In conventional neural operators, the projection layers have been dominated by classical multi-layer perceptrons, which may reduce the precision of solution because they usually suffer from the disadvantages of multi-layer perceptrons, such as lack of interpretability, overfitting, etc. Therefore, to improve the accuracy of the MT forward modeling with neural operators, we integrate the Fourier neural operator with the Kolmogorov-Arnold network (KAN). Specifically, we adopt KAN as the trunk network instead of the classic multi-layer perceptrons to project the resistivity and phase, determined by the branch network-Fourier neural operator, to the desired locations and frequencies. Experimental results demonstrate that the proposed method can achieve high precision in obtaining apparent resistivity and phase at arbitrary frequencies and/or locations with rapid computational speed. 
\end{abstract}
 
\begin{keywords}
Magnetotelluric forward modeling \sep Kolmogorov-Arnold network \sep Neural operator
\end{keywords}

\maketitle 

\printcredits

\doublespacing

\section{Introduction}
Magnetotelluric (MT) has been widely used to illuminate the deep Earth by inverting the sampled electromagnetic field on the Earth surface. MT utilizes natural electromagnetic field induced by solar and ionospheric currents, which interact with the resistivity of the earth to produce secondary fields measurable at the surface. MT has played a crucial role in various applications, including exploration of natural resources \citep{jiang_application_2022, smith_electromagnetic_2014,wang_divergence-free_2024}, geothermal energy assessment \citep{cheng_imaging_2022, gao_three-dimensional_2018}, environmental monitoring \citep{romano_sensitivity_2014}, and characterization of deep geological structures \citep{egbert_fluid_2022, yu_new_2020,yu_three-dimensional_2024}.

The interpretation of MT data is grounded in the inverse problem of determining the resistivity of the Earth from the observed electromagnetic field variations. The classic iterative MT inversion methods (e.g. nonlinear conjugate gradient \citep{newman_three-dimensional_2000} and OCCAM \citep{siripunvaraporn_wsinv3dmt_2009}) consist of forward modeling and model optimization. The essence of forward modeling is to solve the Helmholtz equation for a given conductivity model. The model optimization is to search for the optimal solution by minimizing the mismatch between the observation and the forward modeling electromagnetic field. The efficiency of the MT inversion is usually significantly determined by forward modeling, especially for the high-dimensional MT data inversion. In recent decades, numerical computational methods such as the finite difference method (FDM) \citep{varilsuha_3d_2018}, finite element method \citep{zhu_scalable_2022}, and the finite volume method \citep{guo_modular_2020} have dominated the solution of forward related PDEs of MT. These methods solve the PDEs with physical laws and thus can obtain reasonable solutions. However, their precision largely depends on spatial discretization; finer mesh grids yield higher accuracy but also increase computational costs.

Deep neural networks (DNNs) are powerful for rapid MT forward modeling, which can directly map the conductivity to the observation (e.g. electromagnetic field or apparent resistivity) on the Earth surface. \citet{conway_inverting_2019} trained a DNN that can output apparent resistivity and phase for three-dimensional (3-D) MT forward modeling. \citet{shan_application_2022} applied multitask learning (that is, a DNN has two branches) to predict the apparent resistivity and the phase of the mode $xy$. \citet{wang_three_2024} employed the 3-D Swin Transformer \citep{liu_swin_2021} for 3-D MT forward modeling, predicting the apparent resistivity and phase in different polarization directions. One main disadvantage of these methods is that their efficiency is limited by spatial discretization of the resistivity model, meaning that the trained DNN only works for specific resolutions and frequencies. Thus, it is difficult to achieve satisfactory results on meshes or at frequencies that differ from the training data.

To increase the efficiency of solving PDEs, surrogate models have been developed to approximate solutions, replacing the need for numerical computational methods. Modern machine learning techniques have been used as surrogate models for PDE solvers. Representative models are physics-informed neural networks (PINNs) \citep{raissi_physics-informed_2019} and neural operators (NOs) such as Fourier neural operators (FNOs) \citep{li_fourier_2021} and deep operator networks (DeepONets) \citep{lu_learning_2021}.

The PINNs solve PDEs by modeling the sought solution with a DNN by minimizing the physics-informed loss function through automatic differentiation. \citet{zhang_maxwells_2020} reported that they used PINNs for the simulation of time-domain electromagnetic fields and obtained the electromagnetic fields with a high degree of accuracy, without discretization or interpolation in space or time. Although PINNs are mesh independent, they have limited generalization capabilities, as they require retraining for different conductivities and frequencies, which hinders their use in rapid MT forward modeling.

In contrast, NOs have been proposed to solve PDEs, where NNs map functions to functions \citep{shukla_comprehensive_2024}. FNOs represent a novel deep learning architecture designed to learn mappings between function spaces, especially for solving PDEs. In addition, FNOs are mesh independent and can solve PDEs directly by learning the underlying patterns in the data. At the heart of FNOs is the Fourier layer, which leverages the Fourier transform to convert functions from the spatial domain to the frequency domain. This layer focuses on low frequency that capture the essential features of the functions, applies a linear transformation, and then reverts back to the spatial domain with an inverse Fourier transform. This approach is not only efficient but also aligns well with the global and continuous nature of PDEs. For example, \citet{peng_rapid_2023} proposed a surrogate model for MT forward modeling in the frequency domain by combining FNO with a PDE loss function, eliminating the need for labeled data in network training.

However, FNOs have limitations in MT forward modeling because they require the output resolution to match the input resolution, restricting their ability to obtain responses at arbitrary locations and frequencies. To address this shortcoming, DeepONets introduced the trunk net to map the output of the branch net to the desired solution based on the universal approximation theorem of operators \citep{lu_learning_2021}. \citet{peng_rapid_2022} proposed the extended Fourier neural operator (EFNO) for MT forward modeling in the frequency domain by extending FNO with multilayer perceptron (MLP). The architecture of EFNO is well suited for the demands of responses at arbitrary locations and frequencies for MT forward modeling.

As an alternative to multi-layer perceptrons (MLPs), KANs, as proposed by \citet{liu_kan_2024}, address the shortcomings of MLPs, including lack of interpretability \citep{cranmer_interpretable_2023} and overfitting. Unlike traditional NNs that use fixed activation functions at the nodes, KANs introduce learnable activation functions at the edges that connect the nodes. These activation functions are typically parameterized by certain functions, such as spline functions, providing a high degree of flexibility and adaptability to model complex relationships with potentially fewer parameters. KANs have been explored for solving PDEs and operator learning. \citet{liu_kan_2024} combined PINNs with KANs to solve the 2-D Poisson equation. \citet{abueidda_deepokan_2024} proposed a radial basis functions (RBF)-based KAN operator to solve orthotropic elasticity problems. \citet{shukla_comprehensive_2024} conducted a comprehensive and fair comparison between MLP and KAN for PDE and operator learning.

In this study, we exploit the potential of KANs to improve the accuracy of MT forward modeling by developing a novel approach that takes advantage of FNOs, DeepONets, and KANs. Specifically, based on the basic framework of DeepONets, we utilize FNO as the branch network to handle the frequency information of the resistivity model, and we take the KAN as the trunk net to obtain the apparent resistivity and phase at the desired locations and frequencies. Furthermore, we employ the spectral method to generate the resistivity model with anomalies, rather than simply embedding anomalies within homogeneous half-space underground to evaluate the effectiveness of the proposed method. This paper is organized as follows. In section~\ref{cpt2}, we set up the problem formulation, briefly describe neural operators and KANs, and then provide details about the proposed method. In section~\ref{cpt3}, we demonstrate the effectiveness of the proposed method through numerical experiments. In section~\ref{cpt4}, we provide a discussion of the characteristics of the proposed method. Finally, we draw our conclusions in section~\ref{cpt5}.

\section{Methodology}
\label{cpt2}
In this section, we first establish the MT forward problem with a surrogate model, followed by describing the proposed approach.

\subsection{Problem Formulation}
In the frequency domain, source-free propagation of the natural electric field $\textbf{E}$ and the magnetic field $\textbf{H}$ can be expressed using the simplified Maxwell equations,

\begin{equation}
    \left\{\begin{array}{l}
    \nabla \times \textbf{H}=(\sigma - i \omega \varepsilon) \textbf{E}, \\
    \nabla \times \textbf{E}=i \omega \mu \textbf{H},
    \end{array}\right.
\label{2-1}
\end{equation}

\noindent where $\sigma (S/m)$ represents the conductivity of the subsurface medium, which is the reciprocal of resistivity $\rho (\Omega \cdot m)$, $i$ is the imaginary unit, $\varepsilon(F/m)$ denotes the permittivity, $\omega=2 \pi f$ indicates the angular frequency of the electromagnetic field, and $\mu(H/m)$ denotes the permeability. For MT, it is commonly assumed that the permittivity and permeability of the earth are constants, e.i. $\varepsilon_0=8.85 \times 10^{-12} \mathrm{~F} / \mathrm{m}$ and $\mu_0=4 \pi \times 10^{-7} \mathrm{H} / \mathrm{m}$. Due to $\omega \varepsilon \ll \sigma$, the displacement current can be neglected. Therefore, the electromagnetic field studied by the electromagnetic equations will satisfy the quasi-stationary field assumption within the frequency range of MT, and the equation~\ref{2-1} can be rewritten as

\begin{equation}
    \begin{cases}\nabla^{2} \textbf{E} -k^2 \textbf{E} =0, \\ \nabla^{2} \textbf{H} - k^2 \textbf{H}  =0,\end{cases}
\label{2-2}
\end{equation}

\noindent where $k=\sqrt{-iw\mu \sigma}$. Thus, the number of PDEs to be solved can be reduced to those related to $\textbf{E}$ or $\textbf{H}$, instead of solving all PDEs related to both $\textbf{E}$ and $\textbf{H}$. For 2-D MT forward modeling, the conductivity varies only in the vertical direction $z$ and the horizontal direction $y$; therefore, we can decouple equation~\ref{2-2} into two modes: mode $xy(E_x, H_y, H_z)$ and mode $yx(H_x, E_y, E_z)$ by

\begin{equation}
\left\{\begin{array}{lll}
\nabla \cdot \nabla E_x-k^2 E_x =0, & \text{for mode}\ xy \\ 
\nabla \cdot \nabla H_x  - k^2 H_x =0. & \text{for mode}\ yx
\end{array}\right.
\label{2-3}
\end{equation}

Equation~\ref{2-3} can be solved by cooperating with the appropriate boundary condition, such as the Dirichlet boundary condition,

\begin{equation}
    u_{b c}=g(y, z),
\label{2-4}
\end{equation}

\noindent where $u_{b c}$ presents the electric or magnetic field at the boundary, and $g(y, z)$ denotes the Dirichlet boundary condition at the coordinates $y$ and $z$. Once we obtain $\textbf{E}$ and $\textbf{H}$, we can determine the apparent resistivity $\rho_{xy}$ and $\rho_{yx}$, as well as the phase $\phi_{xy}$ and $\phi_{yx}$, using

\begin{equation}
\begin{aligned}
    & \rho_{x y}=\frac{1}{\omega \mu}\left|\frac{E_x}{H_y}\right|^2, \quad \phi_{x y}=\arctan \left(\frac{\operatorname{Im}\left(E_x / H_y\right)}{\operatorname{Re}\left(E_x / H_y\right)}\right), \\
    & \rho_{y x}=\frac{1}{\omega \mu}\left|\frac{E_y}{H_x}\right|^2, \quad \phi_{y x}=\arctan \left(\frac{\operatorname{Im}\left(E_y / H_x\right)}{\operatorname{Re}\left(E_y / H_x\right)}\right),
\end{aligned}
\label{2-5}
\end{equation}

\noindent where Re denotes the real part and Im denotes the imaginary part, respectively. Additionally, our primary concern is the electromagnetic field on the Earth surface, as this is where we are typically able to deploy instruments for the collection of electromagnetic field data. To obtain the apparent resistivity and phase on the surface, conventional numerical modeling methods, such as FDM, are often used to solve equation~\ref{2-3}, which can be represented as

\begin{equation}
    G_\theta:\left\{\begin{array}{l}
    \sigma(y, z) \\
    \left(y, z, f\right)
    \end{array}\right\} \rightarrow\left\{\begin{array}{l}
    \rho_{x y}\left(y, z, f\right) \\
    \phi_{x y}\left(y, z, f\right) \\
    \rho_{y x}\left(y, z, f\right) \\
    \phi_{y x}\left(y, z, f\right)
    \end{array}\right\},
\label{2-6}
\end{equation}

\noindent where $G_\theta$ indicates the solver with parameter $\theta$ for the desired apparent resistivity and phase. The classical linear MT forward modeling method, such as FDM, usually requires discretizing the conductivity, followed by constructing linear systems for different frequencies. Hence, the computational cost of conventional methods is very high as their performance depends on the number of grids, and they require constructing a separate operator for each frequency. Therefore, it is necessary to develop a surrogate model of $G_\theta$ to improve the efficiency of MT forward modeling.

\subsection{Solving the Problem}
\subsubsection{Neural Operators}
Neural operators have been used to solve PDEs with a modern machine learning technique, which is the solution operator. Representative models are FNOs \citep{li_neural_2020} and DeepONets \citep{lu_learning_2021}. 

\begin{figure*}
\centering
\includegraphics[width=0.7\textwidth]{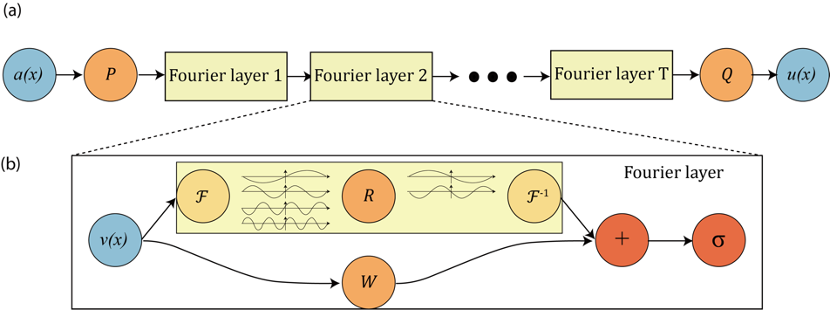}
\caption{(a) The full architecture of FNO \citep{li_fourier_2021}: $P$ is the lifting layer, and $Q$ represents the projecting layer. (b) Fourier layers: $\mathcal{F}$ and $\mathcal{F}^{-1}$ are the Fourier and inverse Fourier transform, respectively, and $W$ denotes linear transform.}
\label{fno}
\end{figure*}

To construct a DNN for solving PDEs, \citet{li_neural_2020} proposed the NO, which can be written as

\begin{equation}
\begin{array}{r}
    v_{j+1}(x)=\alpha\left(W v_j(x)+\int_D \kappa(x, y) v_j(y) d y\right),     j=0,1,2, \ldots, h.
\end{array}
\label{2-7}
\end{equation}

\noindent where $v_{j+1}$ denotes the output of the $j+1$-th layer of the DNN, $\alpha$ represents the activation function, $v_{j}$ indicates the output of the previous layer, and $W$ is the weight. Compared to the standard DNN, the right-hand side of equation~\ref{2-7} includes an additional integral term, where $\kappa$ signifies the integral kernel function that must be learned from the data. By incorporating this integral term, the NO can extract non-local features from the data. Due to the fact that integral computation is time consuming, \citet{li_fourier_2021} further proposed the FNO by the Fourier layer that transforms the integral operation of equation~\ref{2-7} into frequency domain. The Fourier layer can be expressed as

\begin{equation}
    v_{j+1}=\alpha\left(W v_j+\mathcal{F}^{-1}\left(\mathcal{F}(\kappa) \cdot \mathcal{F}\left(v_j\right)\right)\right),
\label{2-8}
\end{equation}

\noindent where $\mathcal{F}$ denotes the Fourier transform and $\mathcal{F}^{-1}$ is the inverse Fourier transform. As shown in Fig.~\ref{fno}, FNO consists of a lifting layer $P$, a Fourier layer, and a projecting layer $Q$. The lifting layer $P$ maps the input $a$ to a high-dimensional channel space. The Fourier layer transforms the input $v$ into frequency and filters the Fourier coefficients, followed by the inverse Fourier transform. The $W$ represents linear transformations, such as those performed by MLPs. The projecting layer $Q$ projects the data to the desired dimension.

Although FNO can share the same parameters regardless of the discretization used for both input and output \citep{kovachki_neural_2024}, it is not suitable for MT forward modeling. This is because FNO typically requires that input and output data have consistent spatial dimensions. This means that if the input data is a multi-dimensional array with a specific resolution, then the output data should also have the same dimensions and resolution. For MT forward modeling, the results of the forward modeling depend not only on the resistivity model but also on the observation frequency and the location of the electromagnetic field observation.

Unlike FNO, DeepONet is not restricted to a specific architecture and can incorporate various types of neural network architectures into its branch and trunk nets, providing greater flexibility. The formulation of DeepONet can be expressed as

\begin{equation}
    G_\theta(\sigma)(y) \approx \sum_{k=1}^p \underbrace{b_k\left(\sigma\left(x_1\right), \sigma\left(x_2\right), \ldots, \sigma\left(x_m\right)\right)}_{\text {branch}} \underbrace{t_k(y)}_{\text {trunk }},
\label{2-9}
\end{equation}

\noindent where $b_k$ represents $k$-th branch net, $t_k$ denotes $k$-th trunk net, and $p$ is the number of branches. In DeepONet, branch and trunk nets facilitate the approximation of infinite-dimensional mappings from inputs to outputs, offering a highly flexible and efficient approach. The branch network provides the necessary input features, while the trunk network maps these features to the output space, allowing DeepONet to handle complex and high-dimensional operator learning tasks \citep{tianping_chen_universal_1995}. This architecture allows DeepONet to generalize well in a range of scenarios, including different domain geometries, input parameters, and initial and boundary conditions \citep{kontolati_learning_2024}.

\subsubsection{KANs}
KANs \citep{liu_kan_2024} is a novel type of neural network architecture inspired by the Kolmogorov-Arnold representation theorem \citep{kolmogorov_representation_1957,braun_constructive_2009}, which posits that any multivariate continuous function can be represented as a superposition of continuous functions of one variable. This foundational theorem underlies the unique structure of KANs, where traditional fixed activation functions are replaced by learnable functions on the network edges, effectively eliminating the need for linear weight matrices.

The core mathematical formula of the Kolmogorov-Arnold representation theorem is given by

\begin{equation}
    f(\mathbf{x}) = \sum_{q=1}^{Q} g_q\left(\sum_{p=1}^{P} \psi_{p,q}(x_p)\right),
\label{2-10}
\end{equation}

\noindent where $f: [0, 1]^d \rightarrow \mathbb{R}$ is a continuous function, $g_q$ and $\psi_{p,q}$ are continuous univariate functions, $P$ and $Q$ are integers that depend on $d$, the number of variables in $\mathbf{x}$.

In KANs, the activation functions are not fixed; instead, they are learned during the training process, allowing for a more flexible and efficient approximation of complex functions. Each weight parameter in a KAN is replaced by a univariate function. In the original implementation \citep{liu_kan_2024}, $\psi(x)$ is defined as a weighted combination of a basis function $b(x)$ and B-splines,

\begin{equation}
    \psi(x)=w_{b}b(x)+w_s\text{spline}(x),
    \label{2-11}
\end{equation}

\noindent where $b(x)$ and the spline$(x)$ are defined as follows.

\begin{equation}
    b(x)=\frac{x}{1+e^{-x}},
\label{2-12}
\end{equation}

\begin{equation}
    \text{spline}(x) = \sum_{i} c_i B_i(x),
\label{2-13}
\end{equation}

\noindent where $w_b$,$w_s$ and $c_i$ are the weights optimized during training, and $B_i(x)$ are the basis functions B-spline defined over a grid. 

Thus, the proposed method is able to improve interpretability, as the learnable functions can be visualized and understood more intuitively. The training process involves optimizing the weights of these splines to minimize the loss function, adjusting the shape of the spline to best fit the training data. KANs offer a promising alternative to traditional MLPs by leveraging the theoretical foundation of the Kolmogorov-Arnold representation theorem, leading to potentially more efficient and interpretable models for approximating complex functions.

\subsubsection{EFKAN}

\begin{figure*}
\centering
\includegraphics[width=0.8\textwidth]{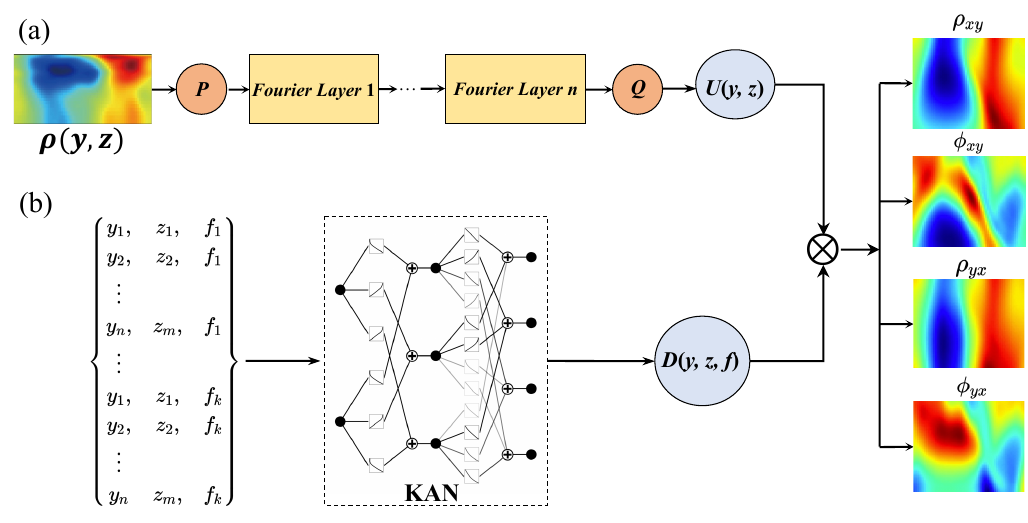}
\caption{The diagram of EFKAN: (a) The branch network consist of lifting layer $P$, Fourier layer, and projecting layer $Q$. The input of the branch network is the resistivity model; (b) the input of the trunk network is the frequencies and coordinates. The output of EFKAN contains of apparent resistivity $\{\rho_{xy}, \rho_{yx}\}$ and phase $\{\phi_{xy}, \phi_{yx}\}$.}
\label{efkan}
\end{figure*}

Considering the characteristics of MT forward modeling, as well as the advantages of FNOs, DeepONets, and KANs, we propose to extend FNO with KAN to improve the accuracy of MT forward modeling; we name it EFKAN (Fig.~\ref{efkan}). Specifically, we employ FNO as the branch net to map the resistivity $\rho(y, z)$ to $U(y,z)$; for the trunk net, we utilize the desired frequencies and coordinates of the electromagnetic field observation as input to KAN to obtain $D(y,z,f)$. Note that $D(y,z,f)$ has the same dimension as $U(y,z)$. We can therefore achieve the desired apparent resistivity and phase ($\rho_{xy}, \phi_{xy}, \rho_{yx}, \phi_{yx}$) by multiplying the matrix of $u_i$ and $d_i$. For the 2-D case, the mathematical expression of MT forward modeling with EFKAN can be written as

\begin{equation}
\begin{aligned}
     G_\theta(\sigma)(y, z, f)  \approx \sum_{k=1}^p \underbrace{b_k\left(\sigma\left(y_1, z_1\right), \sigma\left(y_2, z_2\right), \ldots, \sigma\left(y_m, z_m\right)\right)}_{\text {FNO}} \underbrace{t_k(y, z, f)}_{\text {KAN}}.
\label{2-14}    
\end{aligned}
\end{equation}

In the FNO of EFKAN, we utilize the fast Fourier transformer (FFT), 2-D convolution, and GELU to construct the Fourier layer. In the frequency domain, it is necessary to truncate the high modes to obtain $\mathcal{F}\left(v_t\right) \in \mathbb{C}^{k_{\max} \times d_v}$ because we convolve $v_t \in \mathbb{R}^{n \times d_v}$ with a function that has only $k_{\max}$ Fourier modes. We set $k_{\max}$ to 18. For the 2-D convolution, we define the kernel size to $1\times 1$, and keep the input channels equal to the output channels. Additionally, we place linear layers at the beginning and end of the FNO to determine the dimension of output, and we employ the Gaussian error linear unit (GELU) as the activation function. More details related to FNO can be found in Table~\ref{table-fno}. For the KAN of EFKAN, we use the original KAN implemented in \citep{liu_kan_2024} that adopts the B-spline function as the activation function. The trunk net-KAN has three layers, and the details about each layer are listed in Table~\ref{table-kan}.     

\begin{table}
\renewcommand{\arraystretch}{1.5}
\centering
\caption{FNO architecture overview.}
\label{table-fno}
\begin{tabular}{ccc}
\hline
Layer                                             & Operation                                                                                              & Shape of output                                   \\ \hline
$P$                                                 & Linear                                                                                                           & batch size$\times 64\times 64 \times 1$                                \\ \hline
Fourier layer 1                                   & \begin{tabular}[c]{@{}c@{}}2-D FFT\\ 2-D Convolution\\ GELU \end{tabular} & bacth size$\times 64\times 64\times 32$                               \\ \hline
\begin{tabular}[c]{@{}c@{}}\vdots \end{tabular} & \begin{tabular}[c]{@{}c@{}}\vdots \end{tabular}                                                                & \begin{tabular}[c]{@{}c@{}}\vdots \end{tabular} \\ \hline
Fourier layer 6                                   & \begin{tabular}[c]{@{}c@{}}2-D FFT\\ 2-D Convolution\\ GELU \end{tabular} & bacth size$\times64 \times64 \times 32$                               \\ \hline
$Q1$                                                & Linear                                                                                                           & batch size$\times 64 \times 64 \times 128$                              \\ \hline
Activation                                                & GELU                                                                                                           & batch size$\times 64 \times 64 \times 128$                            
                \\ \hline
$Q2$                                                & Linear                                                                                                           & batch size$\times 64\times 64\times 4$                              \\ \hline
\end{tabular}
\end{table}

\begin{table}
\renewcommand{\arraystretch}{1.5}
\centering
\caption{KAN architecture overview.}
\label{table-kan}
\begin{tabular}{ccccc}
\hline
Layer & Number of neurons & Grid size & Spline order & Grid range  \\ \hline
1     & 2                & 5         & 3            & [-1, 1] \\
2     & 256              & 5         & 3            & [-1, 1] \\
3     & 4096              & 5         & 3            & [-1, 1] \\ \hline
\end{tabular}
\end{table}

\section{Computational Experiments}
\label{cpt3}
In this section, we use synthetic data to evaluate the effectiveness of the proposed method. To make the experiments as close as possible to the field MT forward modeling, we utilize the Gaussian random field (GRF)-based approach to simulate resistivity models for training and testing the EFKAN. We present a series of computational experiments that compare the efficiency and accuracy of EFNO and EFKAN in solving 2-D MT forward problems. Specifically, we focus on using EFKAN to improve the accuracy of solutions for the 2-D Helmholtz equation and explore its applicability to spatial-temporal coordinates and frequencies not included in the training data, as well as its performance on small-scale datasets. All experiments are implemented with the PyTorch platform, and we train and test EFKAN and its competitive approach on a NVIDIA Tesla K80 GPU. It should be noted that the color bars associated with EFKAN errors differ from those associated with EFNO errors in this study. Consequently, when comparing the prediction errors of EFNO and EFKAN, it is essential to take into account the differences between their respective color bars.

\subsection{Data Generation}
To make the conductivity $\sigma$ used for forward modeling more closely resemble the geological structure of the field, we employ the spectral method to generate the conductivity model instead of simply embedding anomalies within a homogeneous half-space underground.

We set the conductivity area of interest to 200 km $\times$ 100 km (width $\times$ height). We discretize the model into 64 grids along the horizontal direction and 64 grids along the vertical direction, with grid intervals that increase in size. Specifically, we divide the space between 0 km and -1 km into 20 grids at fixed intervals, and discretize the area between -1 km and -20 km and the domain between -20 km and -100 km into 20 and 24 grids, respectively, with intervals that increase logarithmically. To ensure that FDM forward modeling fits the boundary conditions of PDEs, we expand the width to 600 km by repeating the conductivity at the margins, extend the depth to -200 km by linearly decaying the conductivity at the bottom, and add an air layer of 600 km $\times$ 200 km. We set the number of grids to 10 for both the expanded area and the air layer. It is worth pointing out that the air layer is required only in mode $xy$.

\begin{figure}
    \centering
    \includegraphics[width=0.5\textwidth]{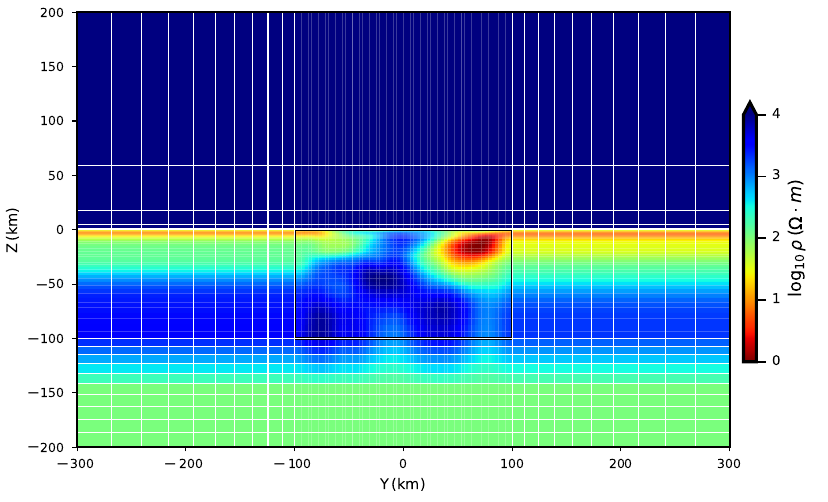}
    \caption{An example of the resistivity model generated through the GRF method.}
\label{conduct_examp}
\end{figure}

To make the synthetic conductivity more realistic, we utilize the spectral method to generate the random field as the conductivity. In the spectral method, the spectrum $P(k)$ is proportional to

\begin{equation}
P(k) \propto|k|^{-\beta / 2},
\label{3-1}
\end{equation}

\noindent where $k$ denotes the wavenumber and $\beta$ represents the scale. Based on common scales of conductivity anomalies found in the earth, we use average conductivity values for five sections of conductivity, with $\beta$ of 3, 4, 5, 6, and 7, respectively. We define the conductivity interval from $10^{-4}$ S/m to 1 S/m, set the conductivity of the air layer to $10^{-9}$ S/m, and the conductivity at the lower boundary to $10^{-2}$ S/m. An example demonstrating the synthetic conductivity model is shown in Fig.~\ref{conduct_examp}.

\begin{table}
\renewcommand{\arraystretch}{1.5}
\centering
\caption{Statistics of training and testing datasets. The frequencies and coordinates of testing dataset A are same with the original training dataset. The number of frequencies and coordinates of testing dataset B are half of those in the original training dataset. The coordinates of C are same with that of the original training dataset, while its frequencies are total different from the original training dataset.}
\label{data_detail}
\begin{tabular}{lcccc}
\hline
Dataset               & Original training dataset                     & Testing dataset A & Testing dataset B & Testing dataset C \\ \hline
Smooth                & \Checkmark  & \Checkmark                                                                    & \Checkmark                                                                    & \Checkmark                                                                    \\
Rectangular anomalies & \ding{56} & \ding{56}                                                                   & \Checkmark                                                                    & \Checkmark                                                                    \\
Number of samples     & 15000                      & 100                                                                                          & 100                                                                                          & 100                                                                                          \\
Number of frequencies & 64                         & 64                                                                                           & 32                                                                                           & 64                                                                                           \\
Number of coordinates & 64                         & 64                                                                                           & 32                                                                                           & 64                                                                                           \\ \hline
\end{tabular}
\end{table}

We uniformly arrange 64 sites within the range of $y = -100$ km to $100$ km along the surface at $z = 0$ km. We employ the FDM to simulate the electric and magnetic fields. We generate 15,000 pairs $\{(\sigma, y, z, f), (\rho_{xy}, \rho_{yx}, \phi_{xy}, \phi_{yx})\}$, comprising the conductivity model, apparent resistivity, and phase, for training EFKAN at 64 frequencies evenly spaced on a logarithmic scale from 0.049 Hz to 10 Hz, referring to this dataset as the original training dataset. Additionally, we generate three datasets to test the performance of EFKAN: (1) Testing dataset A. A has 100 pairs of $\{(\sigma, y, z, f), (\rho_{xy}, \rho_{yx}, \phi_{xy}, \phi_{yx})\}$ with the same frequencies and spatial coordinates as the original training dataset; (2) Testing dataset B. B includes 100 resistivity sections with rectangular anomalies that are unseen in the original training dataset embedded in a smooth background, and the electrical and magnetic fields corresponding to the downsampling frequencies and coordinates of the original training dataset; (3) Testing dataset C. C has 100 resistivity models with rectangular anomalies and the electrical and magnetic fields corresponding to the 64 frequencies (from 0.005 Hz to 12.589 Hz) that are different from those of the original training dataset. The statistics of the training and testing datasets are listed in Table~\ref{data_detail}. Therefore, we will train a network for each of the following datasets: original training dataset, the dataset with downsampling frequencies and coordinates (i.e., 22 frequencies and 22 coordinates), and the downscaling dataset (i.e., 5000 samples). We will then evaluate each model on datasets A, B, and C.

\subsection{Performance on the Original Training Dataset}
\begin{figure}
\centering
\includegraphics[width=0.5\textwidth]{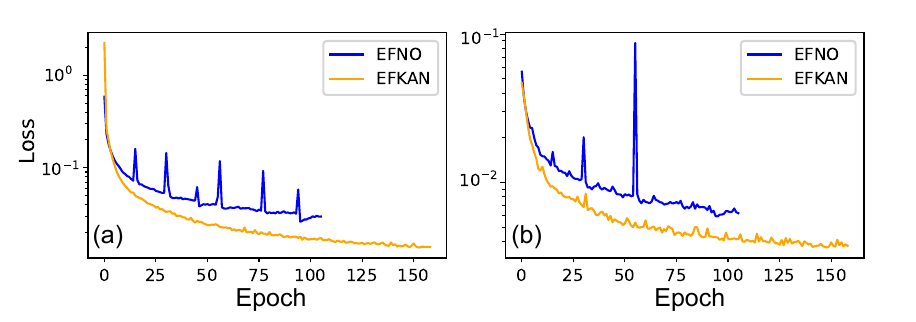}
\caption{(a) The training loss of EFNO and EFKAN on the the original training dataset dataset; (b) The testing loss of EFNO and EFKAN on testing dataset A.}
\label{loss}
\end{figure}

In this section, we evaluate the EFKAN trained by the original training dataset. For both EFNO and EFKAN training, we set the batch size to 50 and the number of training epochs to 200. We optimize both EFNO and EFKAN using the AdamW optimizer with a learning rate of 0.001 by minimizing the relative $\ell_1$-norm loss,

\begin{equation}
\begin{aligned}
        \mathcal{L}=\sum_{i=1}^N \frac{\left\|\rho_{x y, i}-\hat{\rho}_{x y, i}\right\|_1+\left\|\rho_{y x, i}-\hat{\rho}_{y x, i}\right\|_1}{\left\|\rho_{x y, i}\right\|_1+\left\|\rho_{y x, i}\right\|_1+\left\|\phi_{xy, i}\right\|_1+\left\|\phi_{y x, i}\right\|_1}+  \frac{\left\|\phi_{x y, i}-\hat{\phi}_{x y, i}\right\|_1+\left\|\phi_{y x, i}-\hat{\phi}_{yx, i}\right\|_1}{\left\|\rho_{x y, i}\right\|_1+\left\|\rho_{y x, i}\right\|_1+\left\|\phi_{xy, i}\right\|_1+\left\|\phi_{y x, i}\right\|_1},
\label{lossfnc}
\end{aligned}
\end{equation}

\noindent where $\hat{\rho}{xy}, \hat{\rho}{yx}, \hat{\phi}{xy}$, and $\hat{\phi}{yx}$ denote the predicted resistivity and phase for models $xy$ and $yx$, respectively, and $N$ indicates the batch size. Additionally, we apply early stopping for network training, terminating it when the number of epochs with a higher average error than the testing error exceeds 10. The average loss is expressed as

\begin{equation}
  \epsilon=\frac{1}{M}\sum_{i=1}^{K}\mathcal{L}_{i},
\end{equation}

\noindent where $M$ is the number of samples and $K$ indicates the number of iteration per epoch. The loss curves on the original training dataset and testing dataset A are shown in Fig.~\ref{loss}. It can be observed that the loss curves for both EFNO and EFKAN rapidly converge. However, the training and testing loss curves of EFNO fluctuate significantly, and the training of EFNO is terminated at the 106-th epoch. The training and testing loss curves of EFKAN are much smoother and converge to a smaller value than those of EFNO. The quantitative results are listed in Table~\ref{quantive}, which proves that EFKAN can achieve higher precision than EFNO.

\begin{figure}
\centering
\includegraphics[width=0.4\textwidth]{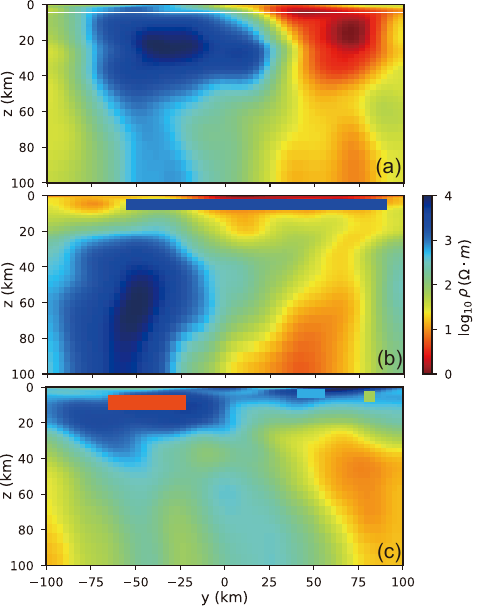}
\caption{Examples for demonstrating the effectiveness of EFKAN: (a) The smooth resistivity model randomly sampled from testing dataset A; (b) The smooth resistivity model with rectangular anomalies sampled from testing dataset B; (c) The smooth resistivity model with rectangular anomalies sampled from testing dataset C.}
\label{t1_models}
\end{figure}

\begin{figure*}
\centering
\includegraphics[width=\textwidth]{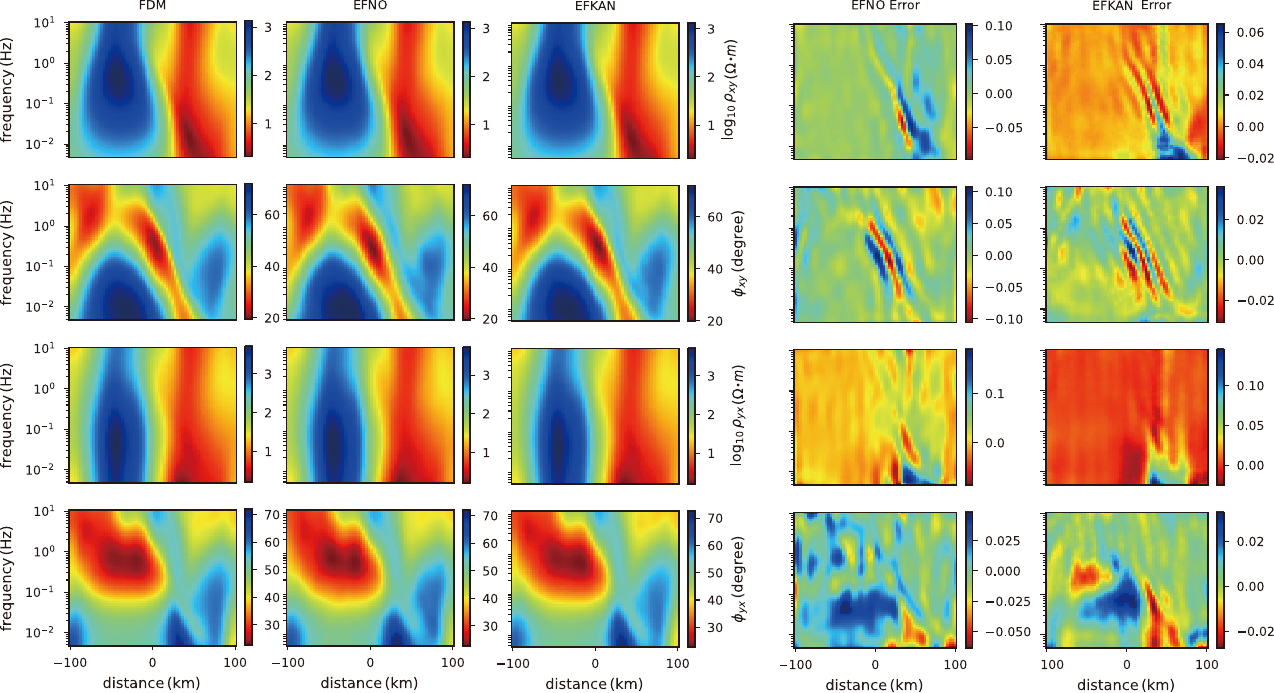}
\caption{Comparison of the ground truth obtained by FDM and the predicted apparent resistivity and phase by EFNO and EFKAN for the smooth resistivity model (Fig.~\ref{t1_models}(a)).}
\label{t1_random_smooth}
\end{figure*}

\begin{figure*}
\centering
\includegraphics[width=\textwidth]{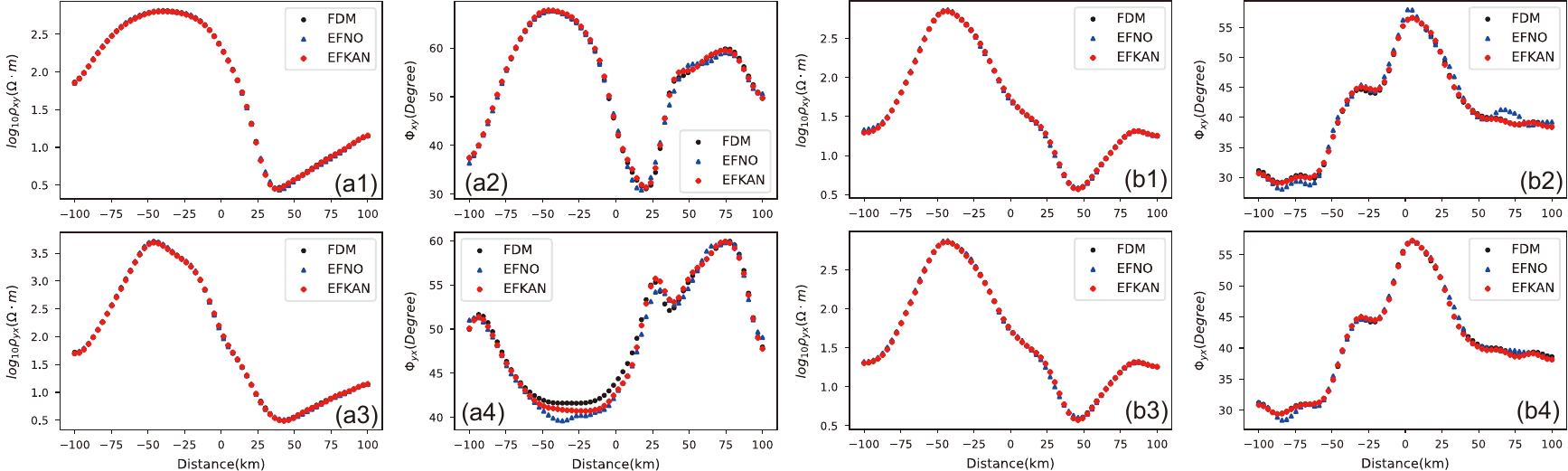}
\caption{The 1-D profiles in Fig.~\ref{t1_random_smooth}: (a1)$\sim$(a4) show the apparent resistivity and phase at 0.049 Hz; (b1)$\sim$(b4) represent the apparent resistivity and phase at 10 Hz.}
\label{t1_1d_smooth}
\end{figure*}

We further randomly select a resistivity mode from each of testing dataset A, B, and C (Fig.~\ref{t1_models}). The predicted apparent resistivity and phase for the smooth resistivity (Fig.~\ref{t1_models}(a)) are shown in Fig.~\ref{t1_random_smooth}, it can be observed that both apparent resistivity and phase predicted by EFNO and EFKAN have high similarities with FDM, and the error by EFKAN is smaller than that of EFNO. The relative $\ell_1$-norm error of $\rho_{xy}, \rho_{yx}, \phi_{xy}$, and $\phi_{yx}$ by EFNO is 0.0081, 0.0150, 0.0178, and 0.0153, respectively. The relative $\ell_1$-norm error of $\rho_{xy}, \rho_{yx}, \phi_{xy}$, and $\phi_{yx}$ by EFKAN is 0.0043, 0.0056, 0.0059, and 0.0093, respectively. The 1-D profiles at 0.049 Hz and 10 Hz (Fig.~\ref{t1_1d_smooth}) further demonstrate that EFKAN has high precision for solving PDEs.

\begin{figure*}
\centering
\includegraphics[width=\textwidth]{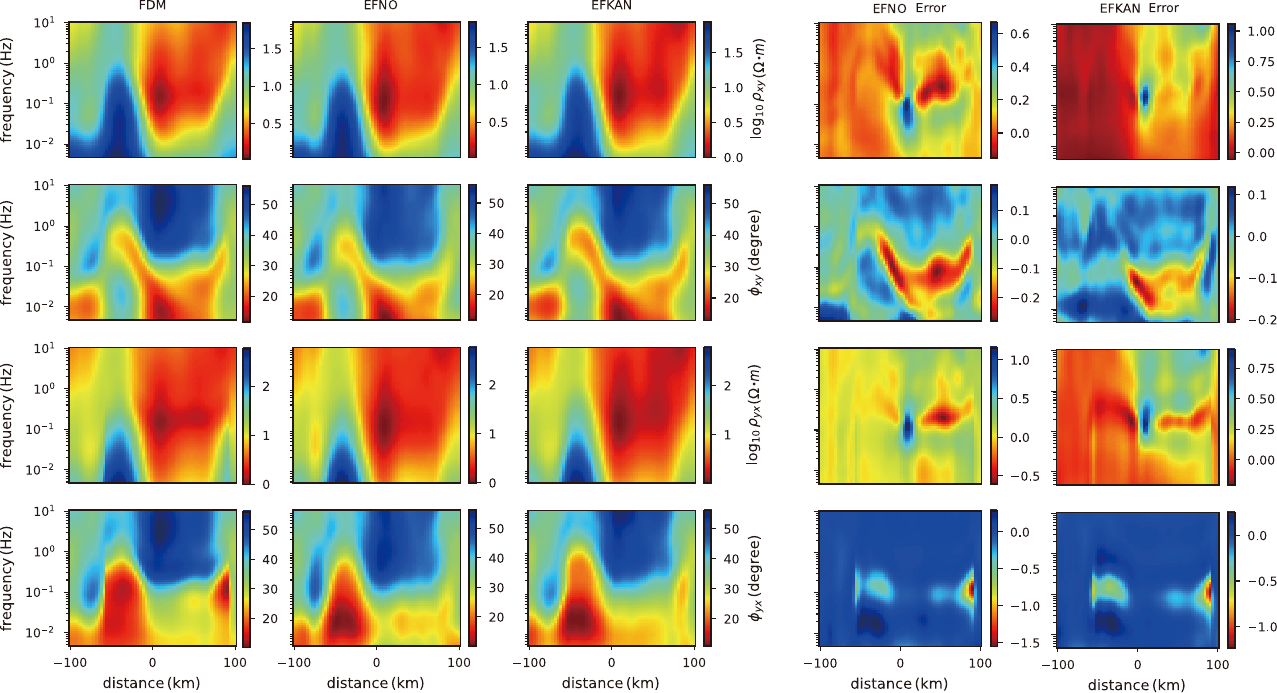}
\caption{Comparison of the ground truth obtained by FDM and the predicted apparent resistivity and phase by EFNO and EFKAN for the smooth resistivity model with rectangular anomalies (Fig.~\ref{t1_models}(b)).}
\label{t1_random_block}
\end{figure*}

\begin{figure*}
\centering
\includegraphics[width=\textwidth]{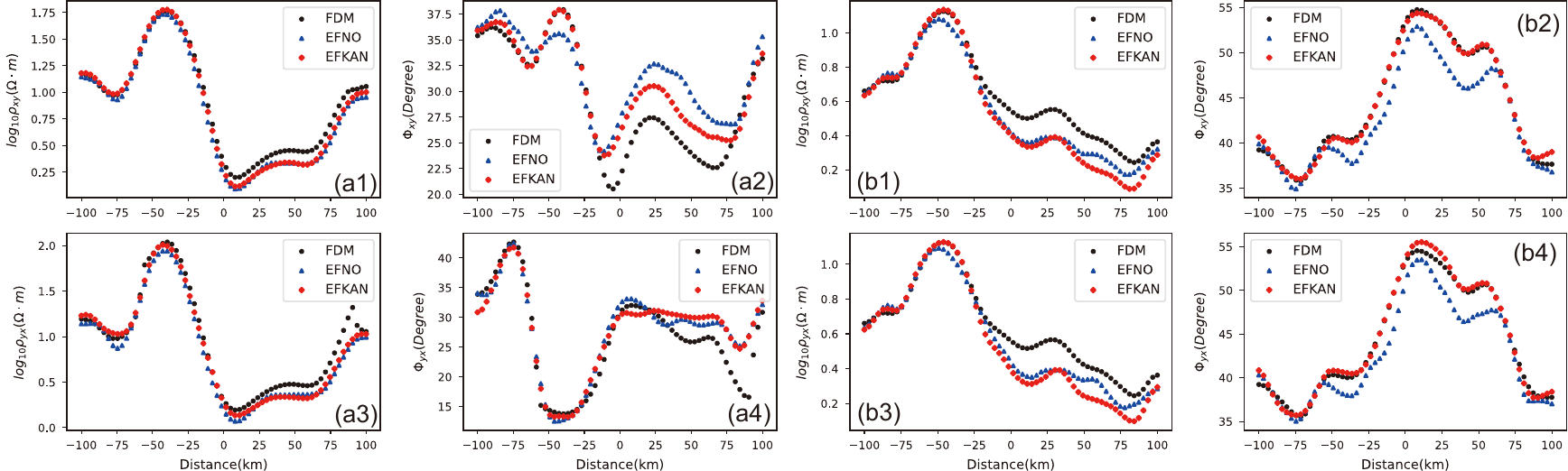}
\caption{The 1-D profiles in Fig.~\ref{t1_random_block}: (a1)$\sim$(a4) show the apparent resistivity and phase at 0.049 Hz; (b1)$\sim$(b4) display the apparent resistivity and phase at 10 Hz.}
\label{t1_1d_block}
\end{figure*}

For the smooth resistivity with rectangular anomalies (Fig.~\ref{t1_models}(b)) from testing dataset B, both EFNO and EFKAN can obtain satisfactory results (Fig.~\ref{t1_random_block}). The relative $\ell_1$-norm error of $\rho_{xy}, \rho_{yx}, \phi_{xy}$, and $\phi_{yx}$ by EFNO is 0.0826, 0.0867, 0.0701, and 0.0852, respectively. The relative $\ell_1$-norm error of $\rho_{xy}, \rho_{yx}, \phi_{xy}$, and $\phi_{yx}$ by EFKAN is 0.0908, 0.0969, 0.0412, and 0.0758, respectively. These errors are higher than those of testing dataset A due to the original training dataset having no rectangular anomalies. As shown in Fig.~\ref{t1_1d_block}, although the predictions from EFNO and EFKAN are essentially the same, EFKAN fits the ground truth (i.e. the results by FDM) better, especially in the phase at 10 Hz (Fig.~\ref{t1_1d_block}(b2) and (b4)).

For the smooth resistivity with rectangular anomalies (Fig.~\ref{t1_models}(c)) from testing dataset C, although the frequencies used for prediction are entirely different from the original training dataset, both EFNO and EFKAN predict the apparent resistivity and phase precisely. The relative $\ell_1$-norm errors for $\rho_{xy}$, $\rho_{yx}$, $\phi_{xy}$, and $\phi_{yx}$ by EFNO are 0.0391, 0.0845, 0.0799, and 0.1349, respectively. The relative $\ell_1$-norm errors for $\rho_{xy}$, $\rho_{yx}$, $\phi_{xy}$, and $\phi_{yx}$ by EFKAN are 0.0451, 0.0877, 0.0666, and 0.1485, respectively. As shown in Fig.~\ref{t1_1d_block_df}, the predictions by EFKAN have high precision with the ground truth in certain areas, such as the apparent resistivity between 0 and 75 km at 12.589 Hz (Fig.~\ref{t1_1d_block_df}(b1) and (b3)).

\begin{table}
\renewcommand{\arraystretch}{1.5}
\centering
\caption{Quantitative results for the original training dataset dataset.}
\label{quantive}
\begin{tabular}{lcccccc}
\hline
\multicolumn{1}{c}{\multirow{2}{*}{Model}} & \multicolumn{2}{c}{Testing dataset A} & \multicolumn{2}{c}{Testing dataset B} & \multicolumn{2}{c}{Testing dataset C} \\ \cline{2-7} 
\multicolumn{1}{c}{}                       & $\epsilon$          & Time (s)          & $\epsilon$          & Time (s)          & $\epsilon$          & Time (s)          \\ \hline
EFNO                                       & 0.0059                               & 0.2302                       & 0.0819                               & 0.2325                       & 0.0785                               & 0.2844                       \\
EFKAN                                      & 0.0037                               & 1.1387                       & 0.0884                               & 1.1141                       & 0.0776                               & 1.2609                       \\ \hline
\end{tabular}
\end{table}

\begin{figure*}
\centering
\includegraphics[width=\textwidth]{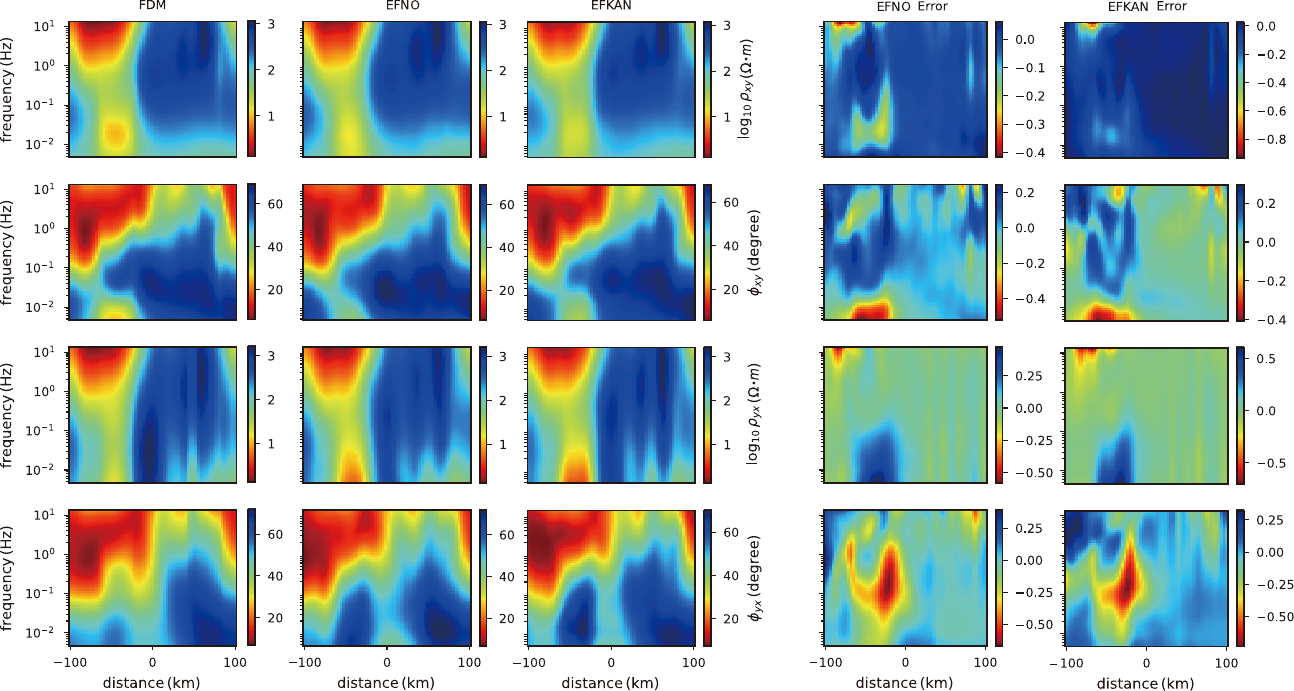}
\caption{Comparison of the ground truth obtained by FDM and the predicted apparent resistivity and phase by EFNO and EFKAN for the smooth resistivity model with rectangular anomalies (Fig.~\ref{t1_models}(c)).}
\label{t1_random_block_df}
\end{figure*}

\begin{figure*}
\centering
\includegraphics[width=\textwidth]{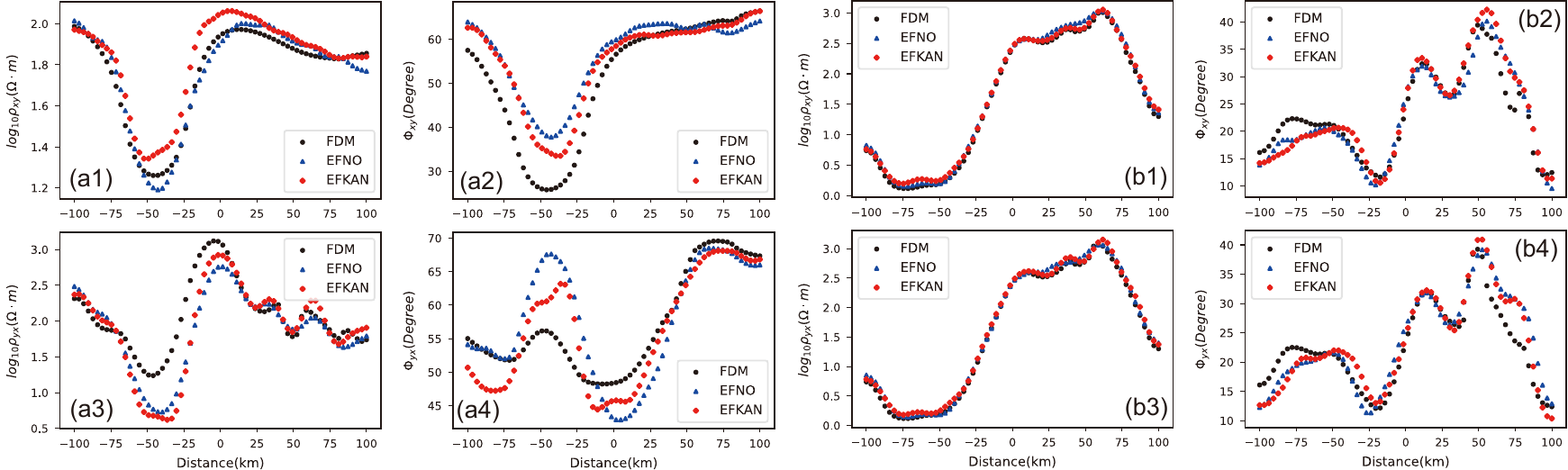}
\caption{The 1-D profiles in Fig.~\ref{t1_random_block_df}: (a1)$\sim$(a4) demonstrate the apparent resistivity and phase at 0.005 Hz; (b1)$\sim$(b4) show the apparent resistivity and phase at 12.589 Hz.}
\label{t1_1d_block_df}
\end{figure*}

\subsection{Performance on the Downsampling Training Dataset}
\begin{figure}
\centering
\includegraphics[width=0.5\textwidth]{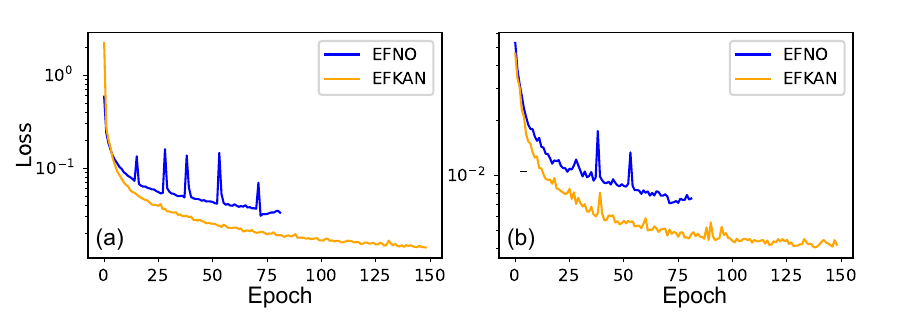}
\caption{(a) The training loss of EFNO and EFKAN on the downsampling original training dataset dataset; (b) The testing loss of EFNO and EFKAN on testing dataset A.}
\label{sp_loss}
\end{figure}

In this section, we downsample both the frequencies and coordinates of the original training dataset by a factor of 3 to train EFNO and EFKAN, aiming to compare their effectiveness in predicting high-resolution data. Specifically, we use the network trained with apparent resistivity and phase at 22 frequencies and 22 coordinates to predict data at 64 frequencies and 64 coordinates. In Fig.~\ref{sp_loss}, we display the training loss in the downsampling original training dataset and the testing loss on testing dataset A. The loss curves of EFNO show fluctuations and stopped at the 82nd epoch. The EFKAN training terminated at the 149-th epoch, and it achieved a smaller error than EFNO. Table~\ref{quantive-sp} clearly shows that EFKAN achieves a smaller average $\ell_1$-norm error on testing dataset A, B, and C, which verifies the effectiveness of EFKAN in predicting high-resolution data. In addition, the quantitative results (Table~\ref{quantive-sp}) demonstrate that EFKAN outperforms EFNO in terms of prediction accuracy, demonstrating its effectiveness for three different kinds of testing data.

\begin{figure}
\centering
\includegraphics[width=0.4\textwidth]{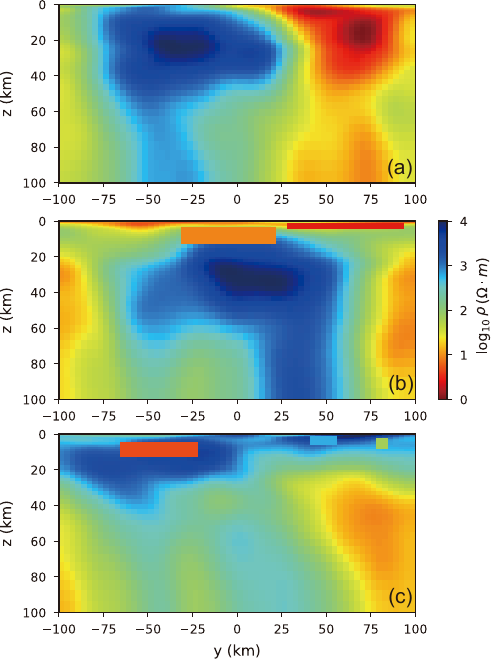}
\caption{Examples for demonstrating the effectiveness of EFKAN trained by the downsampling original training dataset: (a) The smooth resistivity model randomly sampled from testing dataset A; (b) The smooth resistivity model with rectangular anomalies sampled from testing dataset B; (c) The smooth resistivity model with rectangular anomalies sampled from testing dataset C.}
\label{sp_models}
\end{figure}

\begin{figure*}
\centering
\includegraphics[width=\textwidth]{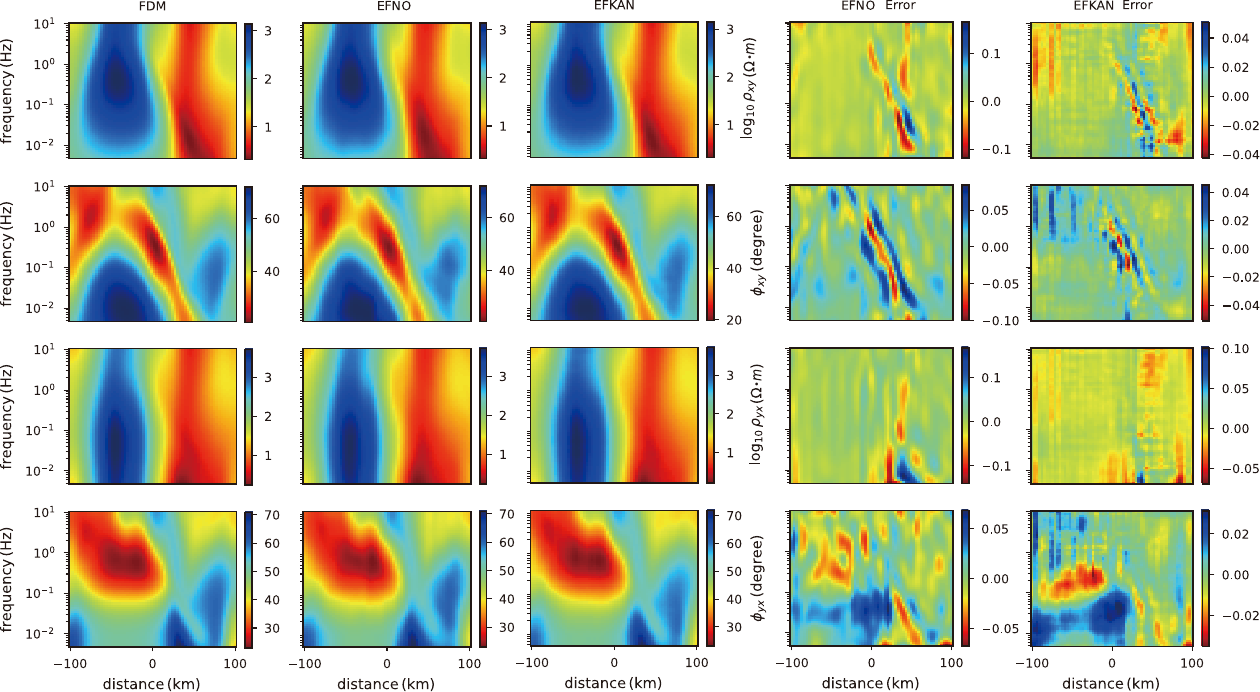}
\caption{Comparison of the ground truth obtained by FDM and the predicted apparent resistivity and phase by EFNO and EFKAN for the smooth resistivity model (Fig.~\ref{sp_models}(a)).}
\label{sp_random_smooth}
\end{figure*}

\begin{figure*}
\centering
\includegraphics[width=\textwidth]{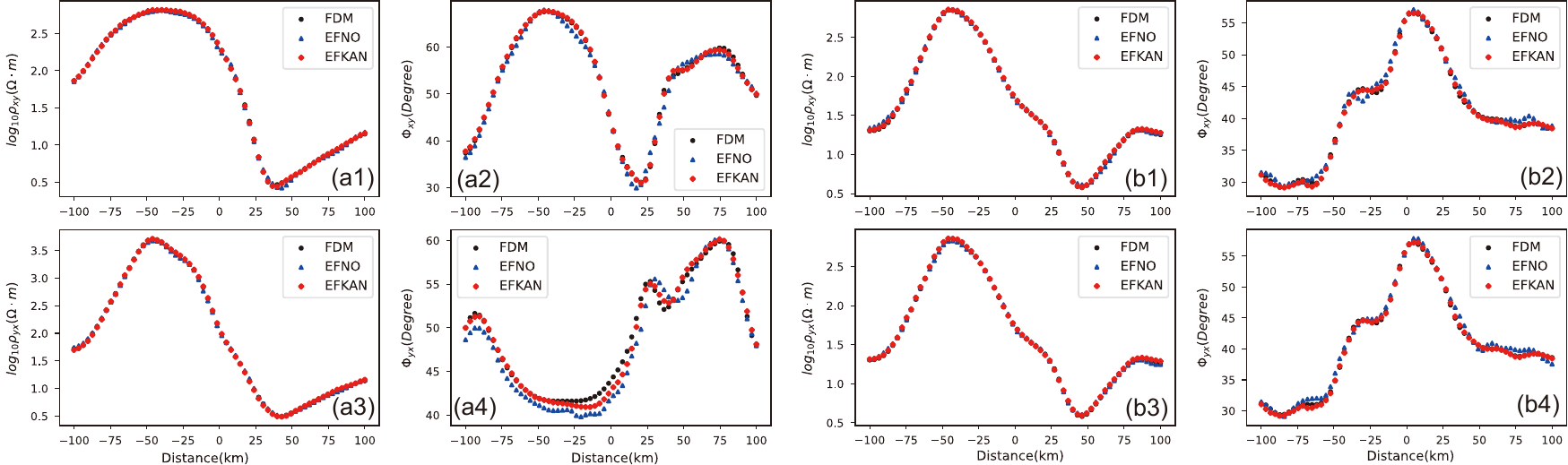}
\caption{The 1-D profiles in from Fig.~\ref{sp_random_smooth}: (a1)$\sim$(a4) display the apparent resistivity and phase at 0.049 Hz; (b1)$\sim$(b4) show the apparent resistivity and phase at 10 Hz.}
\label{sp_1d_smooth}
\end{figure*}

In addition, we randomly sample a resistivity model (Fig.~\ref{sp_models}) from each of testing dataset A, B, and C to show the effectiveness of EFKAN. In Fig.~\ref{sp_random_smooth}, we present the predicted apparent resistivity and phase for smooth resistivity (Fig.~\ref{sp_models}(a)). We can observe that the apparent resistivity and phase predicted by EFNO and EFKAN have high precision, and the $\ell_1$-norm error of EFKAN is lower than that of EFNO, as corroborated by the color bar. The 1-D profiles at 0.049 Hz and 10 Hz (Fig.~\ref{sp_1d_smooth}) further demonstrate that EFKAN can fit the ground truth with higher accuracy than EFNO. The relative $\ell_1$-norm error of $\rho_{xy}, \rho_{yx}, \phi_{xy}$, and $\phi_{yx}$ by EFNO is 0.0107, 0.0123, 0.0190, and 0.0210, respectively. The relative $\ell_1$-norm error of $\rho_{xy}, \rho_{yx}, \phi_{xy}$, and $\phi_{yx}$ by EFKAN is 0.0069, 0.0069, 0.0076, and 0.0110, respectively.

\begin{figure*}
\centering
\includegraphics[width=\textwidth]{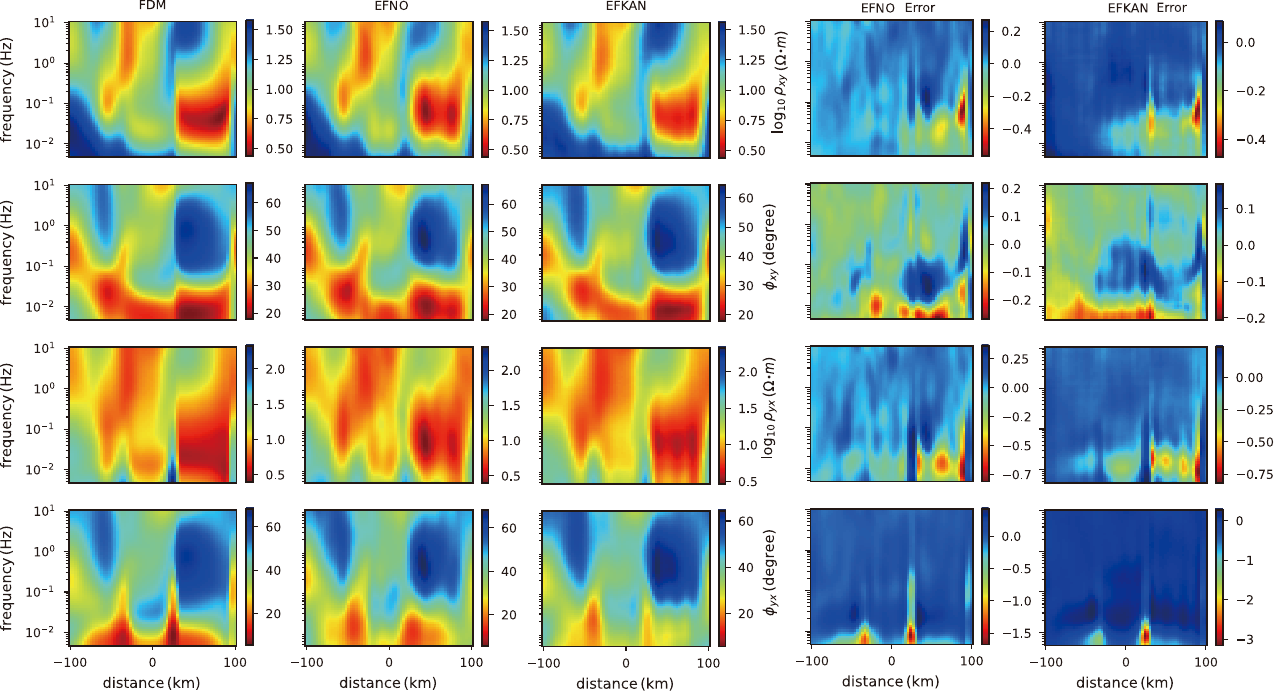}
\caption{Comparison of the ground truth obtained by FDM and the predicted apparent resistivity and phase by EFNO and EFKAN for the smooth resistivity model with rectangular anomalies (Fig.~\ref{sp_models}(b)).}
\label{sp_random_block}
\end{figure*}

\begin{figure*}
\centering
\includegraphics[width=\textwidth]{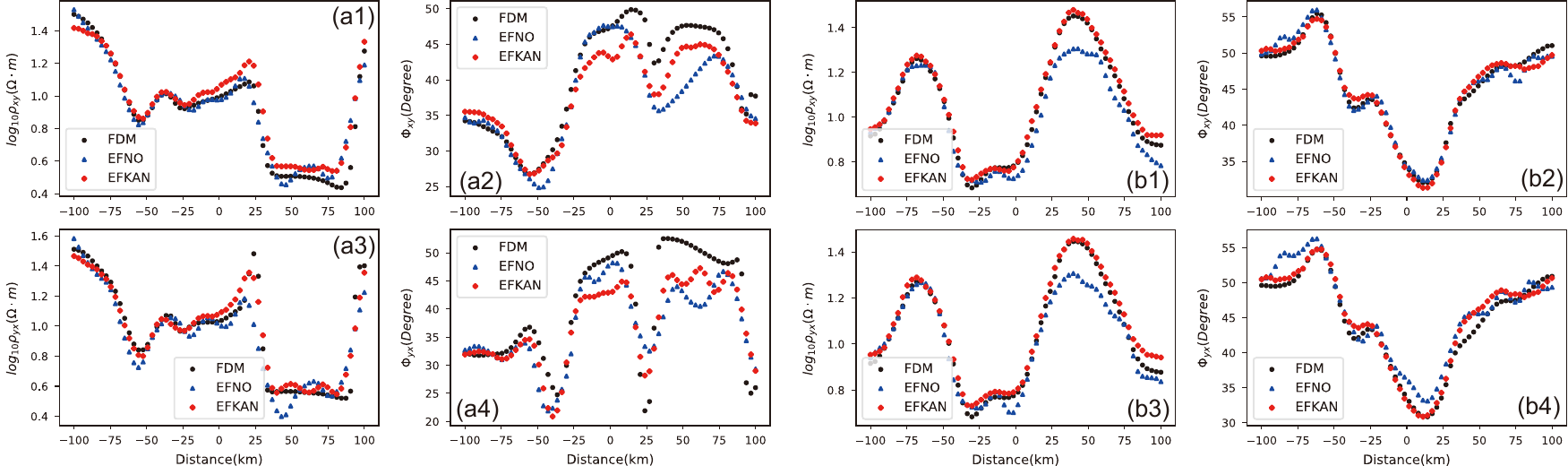}
\caption{The 1-D profiles in Fig.~\ref{sp_random_block}: (a1)$\sim$(a4) show the apparent resistivity and phase at 0.049 Hz; (b1)$\sim$(b4) exhibit the apparent resistivity and phase at 10 Hz.}
\label{sp_1d_block}
\end{figure*}

For the smooth resistivity model with rectangular anomalies (Fig.~\ref{sp_models}(b)) from testing dataset B, although the complexity of the resistivity has increased, both EFNO and EFKAN predict the apparent resistivity and phase with high resolution (Fig.~\ref{sp_random_block}). The relative $\ell_1$-norm error of $\rho_{xy}, \rho_{yx}, \phi_{xy}$, and $\phi_{yx}$ by EFNO is 0.0646, 0.1093, 0.0569, and 0.0969, respectively. The relative $\ell_1$-norm error of $\rho_{xy}, \rho_{yx}, \phi_{xy}$, and $\phi_{yx}$ by EFKAN is 0.0662, 0.0899, 0.0498, and 0.0982, respectively. In the 1-D profiles (Fig.~\ref{sp_1d_block}), we can further observe that the prediction by EFKAN meets the ground truth essentially at 0.049 Hz (Fig.~\ref{sp_1d_block}(a1)$\sim$(a4)) and fits the true apparent resistivity and phase very well at 10 Hz (Fig.~\ref{sp_1d_block}(b1)$\sim$(b4)), outperforming EFNO.

\begin{figure*}
\centering
\includegraphics[width=\textwidth]{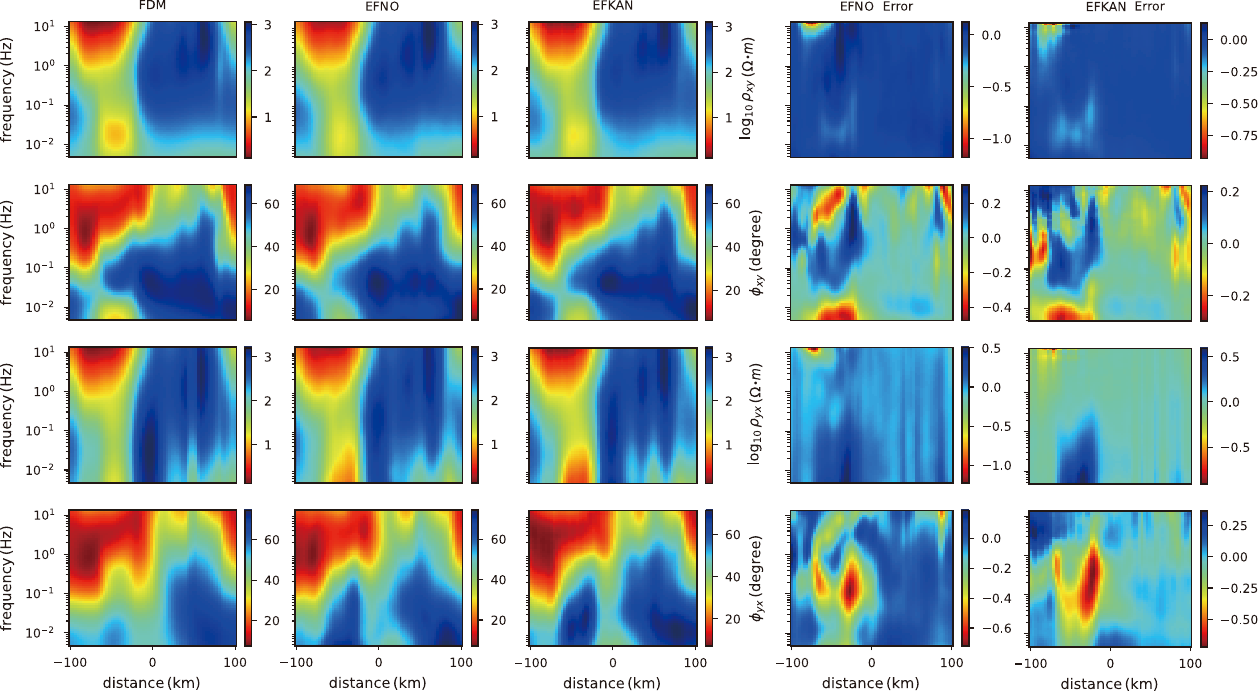}
\caption{Comparison of the ground truth obtained by FDM and the predicted apparent resistivity and phase by EFNO and EFKAN for the smooth resistivity model with rectangular anomalies (Fig.~\ref{sp_models}(c)).}
\label{sp_random_block_df}
\end{figure*}

\begin{figure*}
\centering
\includegraphics[width=\textwidth]{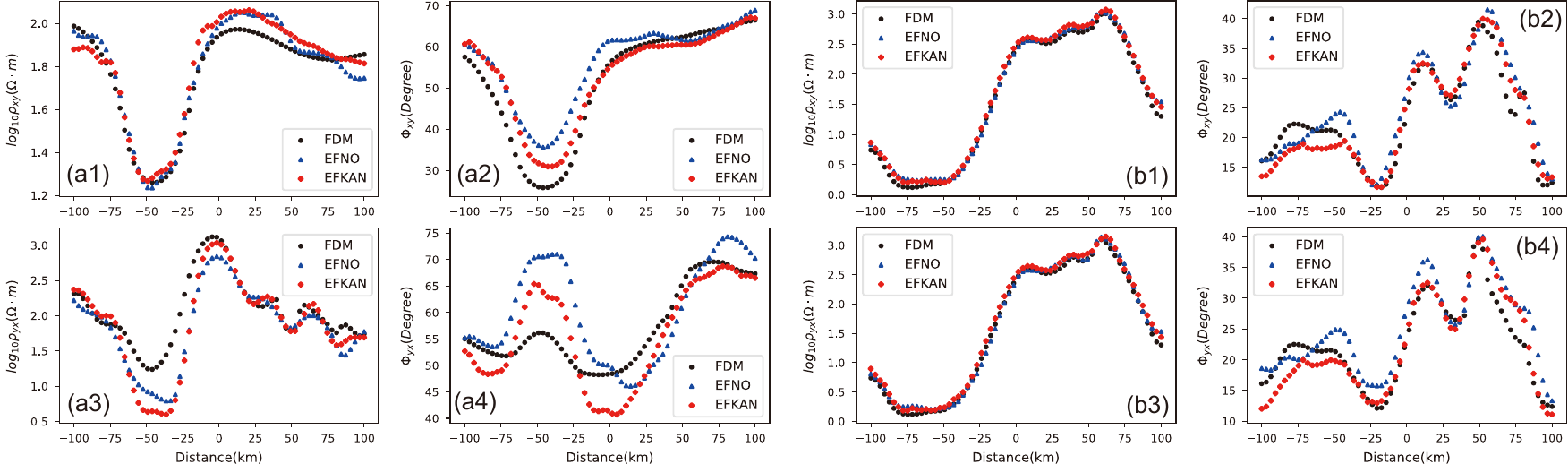}
\caption{The 1-D profiles in Fig.~\ref{sp_random_block_df}: (a1)$\sim$(a4) show the apparent resistivity and phase at 0.005 Hz; (b1)$\sim$(b4) show the apparent resistivity and phase at 12.589 Hz.}
\label{sp_1d_block_df}
\end{figure*}

For the prediction of the frequencies different from those of the training dataset, Fig.~\ref{sp_random_block_df} shows the predictions of the smooth resistivity model with rectangular anomalies (Fig.~\ref{sp_models}(c)) by both EFNO and EFKAN present satisfactory. The $\ell_1$-norm error of EFKAN is lower than that of EFNO. The relative $\ell_1$-norm error of $\rho_{xy}, \rho_{yx}, \phi_{xy}$, and $\phi_{yx}$ of EFNO is 0.0397, 0.0856, 0.0802, and 0.1401, respectively. The relative $\ell_1$-norm error of $\rho_{xy}, \rho_{yx}, \phi_{xy}$, and $\phi_{yx}$ of EFKAN is 0.0415, 0.0904, 0.0604, and 0.1499, respectively. Fig.~\ref{sp_1d_block_df} shows that the solutions of EFNO and EFKAN are similar at 0.005 Hz, while the predictions of EFKAN are more precise than those of EFNO at 12.589 Hz.

\begin{table}
\renewcommand{\arraystretch}{1.5}
\centering
\caption{Quantitative Results for the downsampling original training dataset.}
\label{quantive-sp}
\begin{tabular}{lcccccc}
\hline
\multicolumn{1}{c}{\multirow{2}{*}{Model}} & \multicolumn{2}{c}{Testing dataset A} & \multicolumn{2}{c}{Testing dataset B} & \multicolumn{2}{c}{Testing dataset C} \\ \cline{2-7} 
\multicolumn{1}{c}{}                       & $\epsilon$          & Time (s)          & $\epsilon$          & Time (s)          & $\epsilon$          & Time (s)          \\ \hline
EFNO                                       & 0.0773                               & 0.9933                       & 0.0903                               & 0.9552                       & 0.0795                               & 0.9497                       \\
EFKAN                                      & 0.0041                               & 1.0920                       & 0.0885                               & 1.1158                       & 0.0773                               & 1.1331                       \\ \hline
\end{tabular}
\end{table}

\subsection{Performance on the Downscaling Training Dataset}
\begin{figure}
\centering
\includegraphics[width=0.5\textwidth]{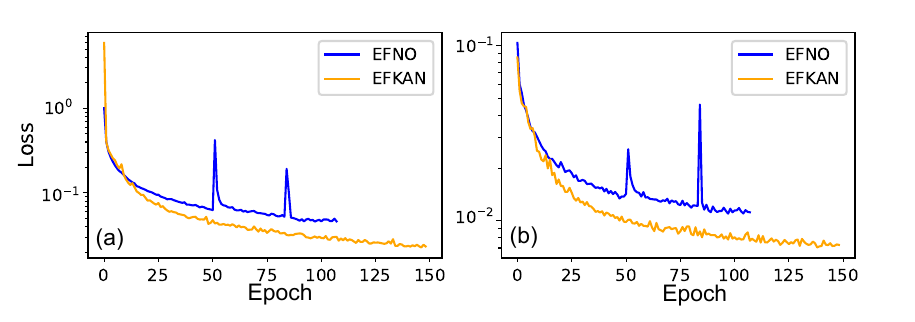}
\caption{(a) The training loss of EFNO and EFKAN on the downscaling original training dataset dataset; (b) The testing loss of EFNO and EFKAN on testing dataset A.}
\label{sd_loss}
\end{figure}

The performance of neural networks is typically limited by the amount of training data. However, the larger the training data, the higher the computational cost. In this section, we reduce the number of samples in the original training dataset to 5000 to assess the generalizability of EFKAN. Fig.~\ref{sd_loss} displays the training loss on the downscaling original training dataset and the testing loss on testing dataset A. The training of EFNO is stopped early at the 108-th epoch, while EFKAN is terminated at the 149-th epoch with a lower loss value. The testing loss curves show that EFKAN achieves a smaller loss than EFNO on testing dataset A. As shown in Table~\ref{quantive-sd}, EFKAN achieves the lowest average $\ell_1$-norm error across the three different testing datasets, demonstrating its strong generalizability.

\begin{figure}
\centering
\includegraphics[width=0.4\textwidth]{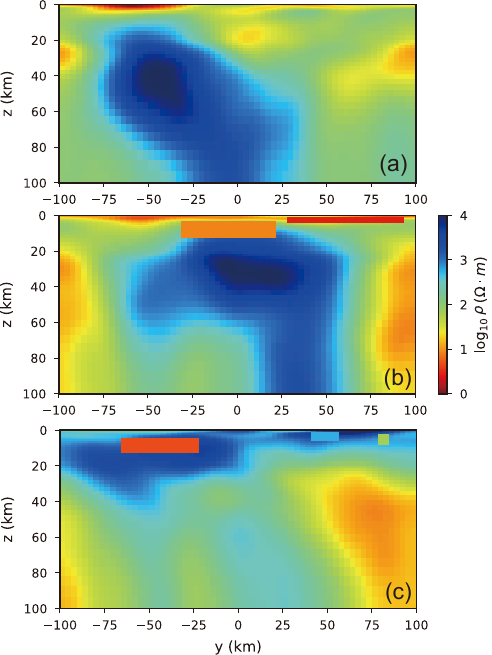}
\caption{Examples for demonstrating the effectiveness of EFKAN trained on the downscaling original training dataset: (a) The smooth resistivity model randomly sampled from testing dataset A; (b) The smooth resistivity model with rectangular anomalies sampled from testing dataset B; (c) The smooth resistivity model with rectangular anomalies sampled from testing dataset C.}
\label{sd_models}
\end{figure}

\begin{figure*}
\centering
\includegraphics[width=\textwidth]{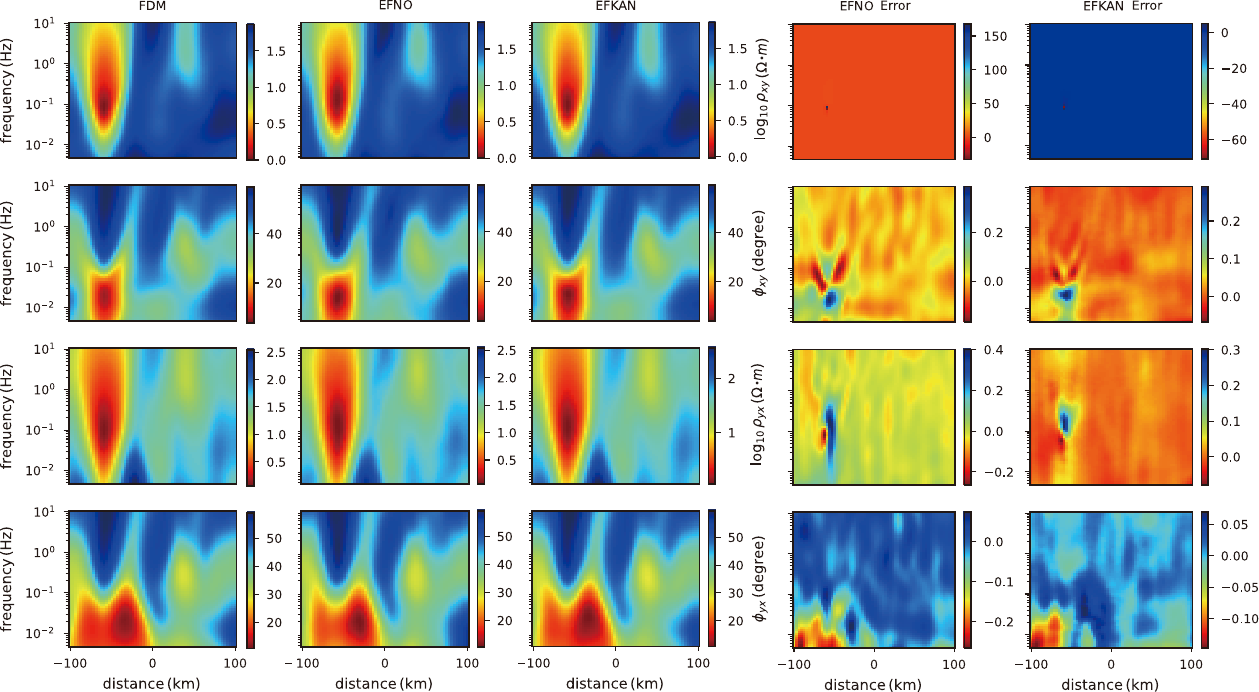}
\caption{Comparison of the ground truth obtained by FDM and the predicted apparent resistivity and phase by EFNO and EFKAN for the smooth resistivity model (Fig.~\ref{sd_models}(a)).}
\label{sd_random_smooth}
\end{figure*}

\begin{figure*}
\centering
\includegraphics[width=\textwidth]{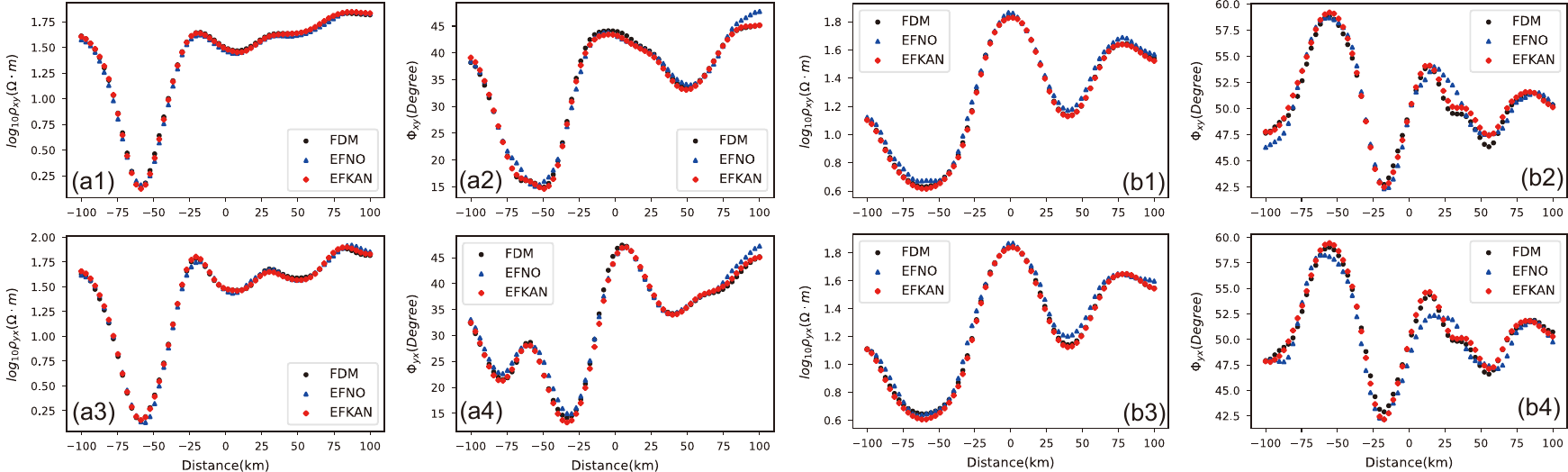}
\caption{The 1-D profiles in Fig.~\ref{sd_random_smooth}: (a1)$\sim$(a4) show the apparent resistivity and phase at 0.049 Hz; (b1)$\sim$(b4) show the apparent resistivity and phase at 10 Hz.}
\label{sd_1d_smooth}
\end{figure*}

We randomly select a resistivity model (Fig.~\ref{sd_models}) from each of testing dataset A, B, and C to show the effectiveness of EFKAN. Fig.~\ref{sd_random_smooth} presents the predicted apparent resistivity and phase for the smooth resistivity (Fig.~\ref{sd_models}(a)), and we can observe that both EFNO and EFKAN obtain solutions with high precision. The $\ell_1$-norm error by EFNO and EFKAN are close to each other except for the error of $\rho_{xy}$, where the amplitude of the error by EFNO is much higher than that of EFKAN. The 1-D profiles at 0.049 Hz and 10 Hz (Fig.~\ref{sd_1d_smooth}) further demonstrate that EFKAN can fit the ground truth better than EFNO. The relative $\ell_1$-norm error of $\rho_{xy}, \rho_{yx}, \phi_{xy}$, and $\phi_{yx}$ by EFNO is 0.0240, 0.0278, 0.0298, and 0.0321, respectively. The relative $\ell_1$-norm error of $\rho_{xy}, \rho_{yx}, \phi_{xy}$, and $\phi_{yx}$ by EFKAN is 0.0088, 0.0123, 0.0163, and 0.0185, respectively.

\begin{figure*}
\centering
\includegraphics[width=\textwidth]{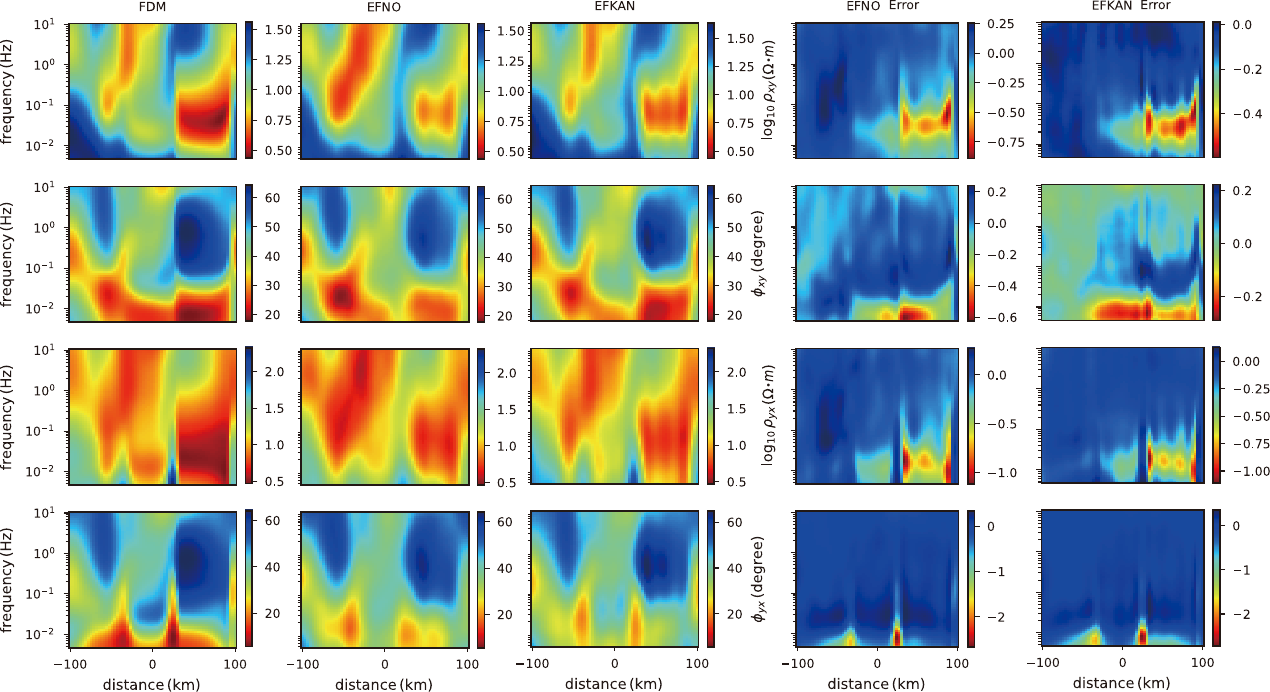}
\caption{Comparison of the ground truth obtained by FDM and the predicted apparent resistivity and phase by EFNO and EFKAN for the smooth resistivity model with rectangular anomalies (Fig.~\ref{sd_models}(b)).}
\label{sd_random_block}
\end{figure*}

\begin{figure*}
\centering
\includegraphics[width=\textwidth]{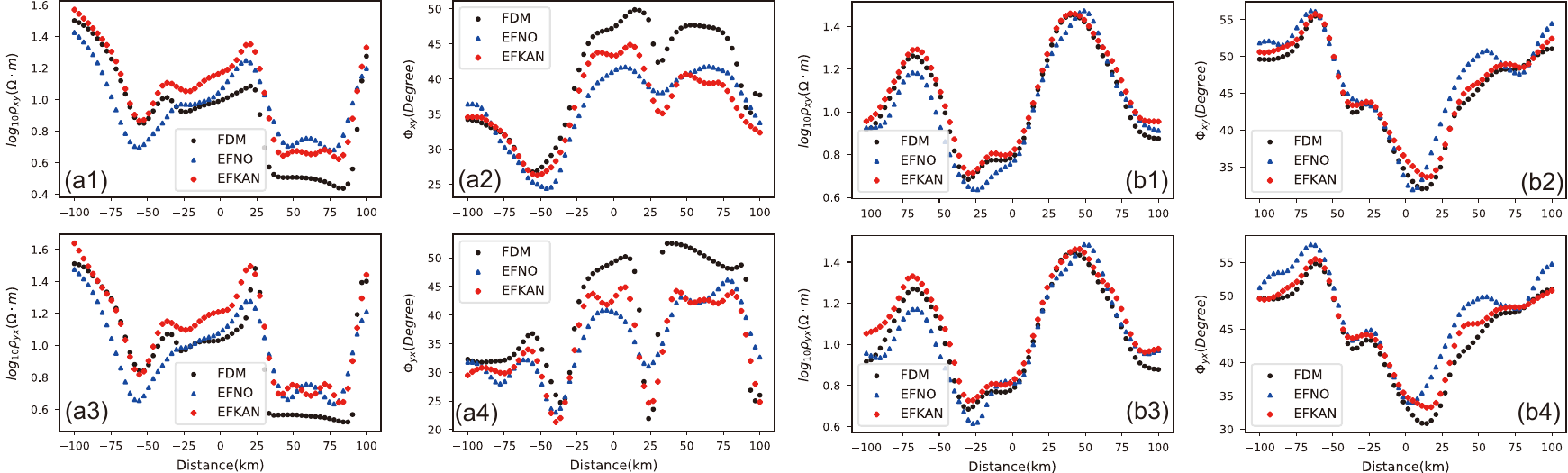}
\caption{The 1-D profiles in Fig.~\ref{sd_random_block}: (a1)$\sim$(a4) show the apparent resistivity and phase at 0.049 Hz; (b1)$\sim$(b4) display the apparent resistivity and phase at 10 Hz.}
\label{sd_1d_block}
\end{figure*}

As shown in Fig.~\ref{sd_random_block}, both EFNO and EFKAN obtain satisfactory results, though the resistivity with rectangular anomalies is not included in the original training dataset. Additionally, the errors are relatively minor and are mainly concentrated in the low frequency. The relative $\ell_1$-norm error of $\rho_{xy}, \rho_{yx}, \phi_{xy}$, and $\phi_{yx}$ by EFNO is 0.1396, 0.1588, 0.0933, and 0.1231, respectively. The relative $\ell_1$-norm error of $\rho_{xy}, \rho_{yx}, \phi_{xy}$, and $\phi_{yx}$ by EFKAN is 0.1161, 0.1425, 0.0702, and 0.1160, respectively. In Fig.~\ref{sd_1d_block}, we present the 1-D profiles, and we can observe that the predictions by EFKAN meet the ground truth well between -100 and -35 km at 0.049 Hz (Fig.~\ref{sd_1d_block}(a1) to (a4)), and it exhibits superior performance to EFNO at 10 Hz (Fig.~\ref{sd_1d_block}(b1) to (b4)).

\begin{figure*}
\centering
\includegraphics[width=\textwidth]{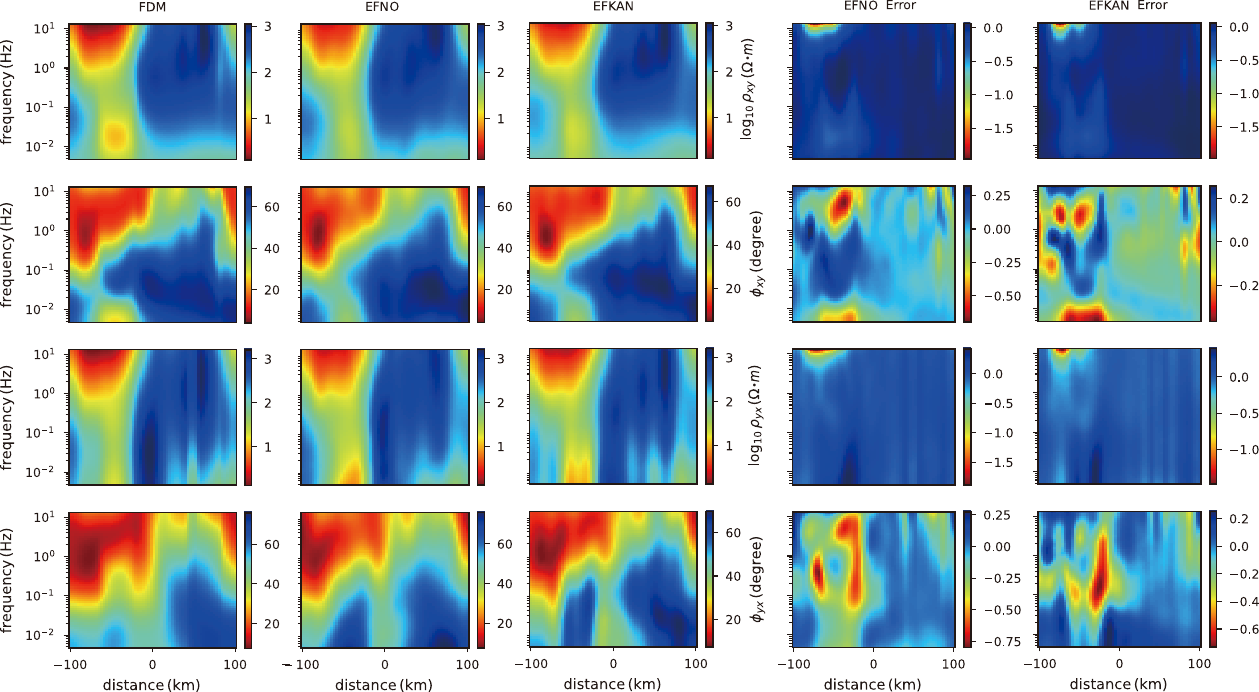}
\caption{Comparison of the ground truth obtained by FDM and the predicted apparent resistivity and phase by EFNO and EFKAN for the smooth resistivity model with rectangular anomalies (Fig.~\ref{sd_models}(c)).}
\label{sd_random_block_df}
\end{figure*}

\begin{figure*}
\centering
\includegraphics[width=\textwidth]{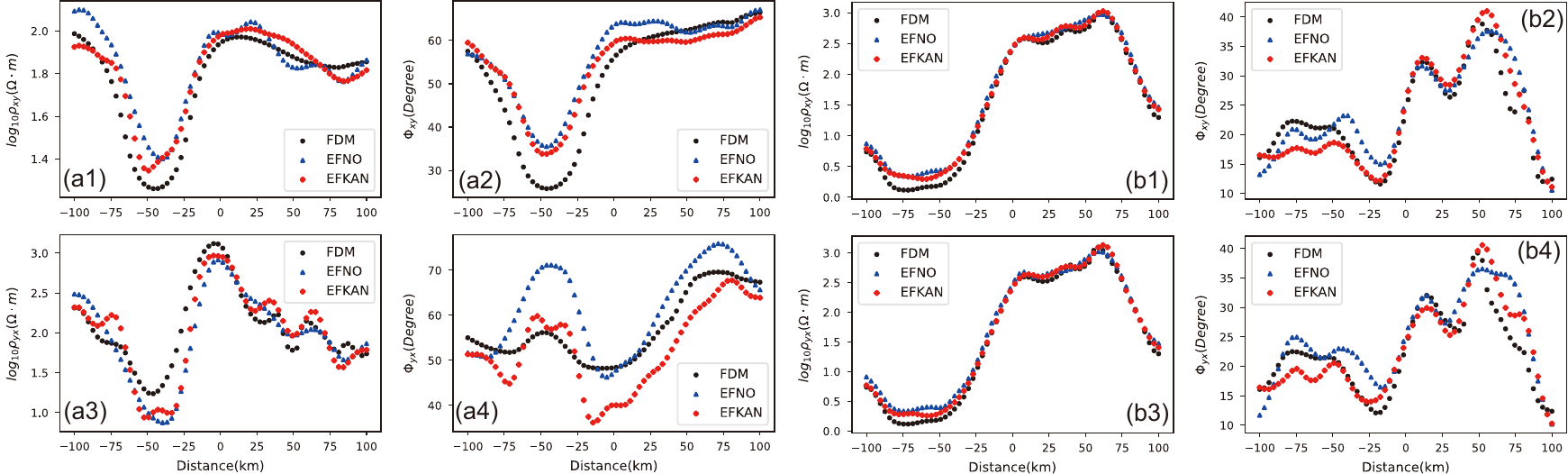}
\caption{The 1-D profiles in Fig.~\ref{sd_random_block_df}: (a1)$\sim$(a4) demonstrate the apparent resistivity and phase at 0.005 Hz;  (b1)$\sim$(b4) represent the apparent resistivity and phase at 12.589 Hz.}
\label{sd_1d_block_df}
\end{figure*}

We also evaluate the performance for the frequencies that are different from the original training dataset. An example is shown in Fig.~\ref{sd_models}, which also includes rectangular anomalies. As shown in Fig.~\ref{sd_random_block_df}, both EFNO and EFKAN are able to provide reasonable apparent resistivity and phase. The errors corresponding to phase are more pronounced than those of apparent resistivity, as well as their distribution. The relative $\ell_1$-norm error of $\rho_{xy}, \rho_{yx}, \phi_{xy}$, and $\phi_{yx}$ by EFNO is 0.0844, 0.0797, 0.0911, and 0.1402, respectively. The relative $\ell_1$-norm error of $\rho_{xy}, \rho_{yx}, \phi_{xy}$, and $\phi_{yx}$ by EFKAN is 0.0529, 0.0767, 0.0683, and 0.1403, respectively. From the 1-D profiles (Fig.~\ref{sd_1d_block_df}), it can be observed that both EFNO and EFKAN can essentially fit the apparent resistivity and phase at 0.005 Hz as well as the phase at 12.589 Hz. Furthermore, the apparent resistivity predicted by the two methods is very close to the ground truth, yet EFKAN is more accurate than EFNO at 12.589 Hz (Fig.~\ref{sd_1d_block_df}(b1) and (b3)).

\begin{table}
\renewcommand{\arraystretch}{1.5}
\centering
\caption{Quantitative Results for the downscaling original training dataset}
\label{quantive-sd}
\begin{tabular}{lcccccc}
\hline
\multicolumn{1}{c}{\multirow{2}{*}{Model}} & \multicolumn{2}{c}{Testing dataset A} & \multicolumn{2}{c}{Testing dataset B} & \multicolumn{2}{c}{Testing dataset C} \\ \cline{2-7} 
\multicolumn{1}{c}{}                       & $\epsilon$          & Time (s)          & $\epsilon$          & Time (s)          & $\epsilon$          & Time (s)          \\ \hline
EFNO                                       & 0.0110                               & 0.9537                       & 0.0949                               & 0.9694                       & 0.0844                               & 0.9742                       \\
EFKAN                                      & 0.0070                               & 1.1269                       & 0.0932                               & 1.1053                       & 0.0825                               & 1.1407                       \\ \hline
\end{tabular}
\end{table}

\subsection{Computational Cost}
From tables~\ref{quantive}, \ref{quantive-sp}, and \ref{quantive-sd}, it is evident that EFKAN requires approximately 0.01 seconds to predict the electromagnetic field for a single resistivity model. In contrast, the traditional and widely-used numerical algorithm FDM takes about 10 seconds per resistivity model for forward modeling. This indicates that the computational speed of EFKAN is nearly 1000 times faster than that of FDM. Additionally, EFKAN consumes slightly more time per resistivity model compared to EFNO, with training time per epoch being 93 seconds and 114 seconds for EFNO and EFKAN, respectively. This discrepancy may arise from the fact that the KAN employs a separate function for each variable in the input, and the standard B-spline functions utilized in the KAN are not optimized for parallel computing on GPUs. To enhance computational efficiency, one could consider replacing the B-spline function with alternative univariate functions, such as radial basis functions (RBFs) \citep{li_kolmogorov-arnold_2024}, wavelets \citep{bozorgasl_wav-kan_2024}, and Jacobi polynomials \citep{ss_chebyshev_2024}.

\section{Discussion}
\label{cpt4}
The experimental results prove that EFKAN can achieve higher accuracy in apparent resistivity and phase measurements compared to the EFNO \citep{peng_rapid_2022} that powered by FNO and MLP. We believe that this is due to the advantage of KAN for science. Based on the Kolmogorov-Arnold representation theorem, KAN has no linear weights at all–every weight parameter is replaced by a univariate function \citep{liu_kan_2024}, such as B-spline functions. Due to the locality and the adjustable number of grids in the univariate function, KANs can achieve a certain degree of dynamic network architecture and continuous learning \citep{yu_kan_2024}. Reference \citep{yu_kan_2024} proved that KAN significantly outperforms MLP in symbolic formula representation, while MLP generally outperforms KAN in machine learning, computer vision, natural language processing, and audio processing. Therefore, KANs have demonstrated promising performance in many science tasks including Poisson equation, knot theory, and Anderson localization.

Furthermore, we believe that KAN can endow EFKAN with a certain degree of interpretability. The Kolmogorov-Arnold representation theorem posits that any multivariate continuous function can be represented as a superposition of continuous functions of a single variable. Thus, it is feasible to visualize and understand these learnable activation functions, allowing for a more intuitive grasp of how inputs are transformed into output; examples are shown in Fig.~\ref{efkan}. This characterization not only overcomes the limitations of fixed activation functions in terms of interpretability and training efficiency but also aligns well with the global and continuous nature of the problems KAN are designed to solve, such as partial differential equations and operator learning tasks. The parameterization of these functions and their learnable nature make KAN more transparent and understandable, compared to the \emph{black-box} nature of traditional MLPs.

From Fig.~\ref{loss}, Fig.~\ref{sp_loss}, and Fig.~\ref{sd_loss}, we can observe that EFKAN training converging rapidly and stably, while EFNO training is fluctuating and stopped early. That proves that the combination of KAN and FNO does not introduce new issues in parameter tuning, and it can improve training stability and convergence. The computational intensity of forward modeling significantly constrains the efficiency of MT inversion. The iterative inversion algorithm, such as the nonlinear conjugate gradient \citep{newman_three-dimensional_2000}, can obtain the inversion result with high resolution by minimizing the misfit between the observed and simulated electromagnetic field. However, the high computation costs required for MT forward modeling restrict exploration of the model parameter space and limit the number of feasible iterations \citep{bozdag_global_nodate}. EFKAN shows two significant advantages for further artificial intelligence-driven MT inversion: (1) By significantly increasing the forward modeling speed to reduce the computational cost of inversion, enabling more extensive searches of the model parameter space within a limited time; (2) By improving the accuracy of forward modeling to optimize search directions within the model parameter space, enhancing the precision of inverted resistivity model. Thus, we believe that EFKAN will improve the efficiency of traditional MT inversion methods, and it can provide a new tool to further enhance the generalization capability of conventional deep learning-based MT inversion, such as the CNN-based MT inversion approach \citep{yu_two-dimensional_2025}.

The generalization capability of deep learning is largely based on the quality and amount of training datasets. To obtain high-quality training data, we use GRF to generate the training data to reflect the essential characteristics of real conductivity, though it may not fully capture all the complexities present in real-world resistivity distributions, such as sharp discontinuities or fault-like structures. This is important for real applications because it is very difficult or expensive to obtain a large amount of real data to train neural networks. From the experimental results, we can see that EFKAN can achieve good performance on the unseen data (e.g., the rectangular anomalies and the frequencies not present in the training dataset) and satisfactory predictions. Therefore, we consider that the proposed method can provide flexibility enabling researchers to customize their own training sets and rapidly deploy the neural networks for specific MT forward modeling tasks.       

In addition, we can observe that the testing loss of both EFNO and ENKAN is lower than the training loss (Fig.~\ref{loss}, Fig.~\ref{sp_loss}, and Fig.~\ref{sd_loss}), which is contrary to that in many deep learning tasks. By analyzing the training and testing dataset, we speculate that this is due to the fact that the number of samples in the test dataset is significantly fewer than the number of samples in the training dataset. Specifically, the original training dataset contains 15000 samples, which is 150 times the number of samples of testing dataset A, B and C. Therefore, the accumulated error on the dataset with more samples may be greater than the accumulated error on the dataset with fewer samples. It is worth mentioning that this phenomenon has also occurred in our previous work \citep{peng_rapid_2022}. We also find this phenomenon existing in the Fashion MNIST classification, which consists of a training set of 60000 samples and a test set of 10000 samples. It can be observed that regardless of whether an MLP or CNN is used, the validation loss (that is, the testing loss in our experiments) is lower than the training loss (\url{https://nvsyashwanth.github.io/machinelearningmaster/fashion-mnist/}).

\section{Conclusion}
\label{cpt5}
In this study, we present the development of a new neural operator for the rapid forward modeling of MT by combining FNO with KAN. We utilize FNO as the branch network to transform the resistivity model to the apparent resistivity and phase in the frequency domain followed by using KAN to map these apparent resistivity and phase at the desired frequencies and/or locations. In addition, we adopt the spectral method to generate smooth and stochastically varying resistivity models as the training data, rather than embedding anomalies within a homogeneous half-space underground. To evaluate the generalization capability of the proposed method, we implement several challenging experiments, including predicting the high-resolution solution using the model trained on low-resolution data and the solution for the unseen data. Experimental results demonstrate that the proposed method has good generalization capability for these challenging tasks, which can achieve high accuracy and is much faster than the traditional numerical computational algorithm in terms of computational speed, suggesting that EFKAN could serve as a potential surrogate model for rapid MT forward modeling. 

\section{Acknowledgments}
This work was supported in part by the National Natural Science Foundation of China under Grant 42474103 and Grant 61906170, in part by the Food Science and Engineering, the Most Important Discipline of Zhejiang Province under Grant ZCLY24F0301, in part by the Basic Public Welfare Research Program of Zhejiang Province under Grant LGF21F020023, in part by the Natural Science Foundation of Ningbo Municipality under Grant 2022Z233, Grant 2021Z050, Grant 2022S002 and 2023J403.

\newpage

\textbf{Code availability}
The source codes are available for download at the link: https://github.com/linfengyu77/EFKAN.

\textbf{Declaration of competing interest}
The authors declare that they have no known competing financial interests or personal relationships that could have appeared to influence the work reported in this paper.

\textbf{Data availability}
No data was used for the research described in the article.

\bibliographystyle{cas-model2-names}
\bibliography{ref} 

\begin{thebibliography}{40}
\expandafter\ifx\csname natexlab\endcsname\relax\def\natexlab#1{#1}\fi
\providecommand{\url}[1]{\texttt{#1}}
\providecommand{\href}[2]{#2}
\providecommand{\path}[1]{#1}
\providecommand{\DOIprefix}{doi:}
\providecommand{\ArXivprefix}{arXiv:}
\providecommand{\URLprefix}{URL: }
\providecommand{\Pubmedprefix}{pmid:}
\providecommand{\doi}[1]{\href{http://dx.doi.org/#1}{\path{#1}}}
\providecommand{\Pubmed}[1]{\href{pmid:#1}{\path{#1}}}
\providecommand{\bibinfo}[2]{#2}
\ifx\xfnm\relax \def\xfnm[#1]{\unskip,\space#1}\fi
\bibitem[{Abueidda et~al.(2024)Abueidda, Pantidis and Mobasher}]{abueidda_deepokan_2024}
\bibinfo{author}{Abueidda, D.W.}, \bibinfo{author}{Pantidis, P.}, \bibinfo{author}{Mobasher, M.E.}, \bibinfo{year}{2024}.
\newblock \bibinfo{title}{{DeepOKAN}: {Deep} {Operator} {Network} {Based} on {Kolmogorov} {Arnold} {Networks} for {Mechanics} {Problems}}.
\newblock \URLprefix \url{http://arxiv.org/abs/2405.19143}, \DOIprefix\doi{10.48550/arXiv.2405.19143}. \bibinfo{note}{arXiv:2405.19143 [cs]}.
\bibitem[{Bozdağ et~al.()Bozdağ, Peter, Lefebvre, Komatitsch, Tromp, Hill, Podhorszki and Pugmire}]{bozdag_global_nodate}
\bibinfo{author}{Bozdağ, E.}, \bibinfo{author}{Peter, D.}, \bibinfo{author}{Lefebvre, M.}, \bibinfo{author}{Komatitsch, D.}, \bibinfo{author}{Tromp, J.}, \bibinfo{author}{Hill, J.}, \bibinfo{author}{Podhorszki, N.}, \bibinfo{author}{Pugmire, D.}, .
\newblock \bibinfo{title}{Global adjoint tomography: first-generation model} \URLprefix \url{https://dx.doi.org/10.1093/gji/ggw356}.
\bibitem[{Bozorgasl and Chen(2024)}]{bozorgasl_wav-kan_2024}
\bibinfo{author}{Bozorgasl, Z.}, \bibinfo{author}{Chen, H.}, \bibinfo{year}{2024}.
\newblock \bibinfo{title}{Wav-{KAN}: {Wavelet} {Kolmogorov}-{Arnold} {Networks}}.
\newblock \URLprefix \url{http://arxiv.org/abs/2405.12832}. \bibinfo{note}{arXiv:2405.12832 [cs, eess, stat]}.
\bibitem[{Braun and Griebel(2009)}]{braun_constructive_2009}
\bibinfo{author}{Braun, J.}, \bibinfo{author}{Griebel, M.}, \bibinfo{year}{2009}.
\newblock \bibinfo{title}{On a {Constructive} {Proof} of {Kolmogorov}’s {Superposition} {Theorem}}.
\newblock \bibinfo{journal}{Constructive Approximation} \bibinfo{volume}{30}, \bibinfo{pages}{653--675}.
\newblock \URLprefix \url{https://doi.org/10.1007/s00365-009-9054-2}, \DOIprefix\doi{10.1007/s00365-009-9054-2}.
\bibitem[{Cheng et~al.(2022)Cheng, Pang, Kong, Chen and Wang}]{cheng_imaging_2022}
\bibinfo{author}{Cheng, Y.}, \bibinfo{author}{Pang, Z.}, \bibinfo{author}{Kong, Y.}, \bibinfo{author}{Chen, X.}, \bibinfo{author}{Wang, G.}, \bibinfo{year}{2022}.
\newblock \bibinfo{title}{Imaging the heat source of the {Kangding} high-temperature geothermal system on the {Xianshuihe} fault by magnetotelluric survey}.
\newblock \bibinfo{journal}{Geothermics} \bibinfo{volume}{102}, \bibinfo{pages}{102386}.
\newblock \URLprefix \url{https://www.sciencedirect.com/science/article/pii/S0375650522000384}, \DOIprefix\doi{10.1016/j.geothermics.2022.102386}.
\bibitem[{Conway et~al.(2019)Conway, Alexander, King, Heinson and Kee}]{conway_inverting_2019}
\bibinfo{author}{Conway, D.}, \bibinfo{author}{Alexander, B.}, \bibinfo{author}{King, M.}, \bibinfo{author}{Heinson, G.}, \bibinfo{author}{Kee, Y.}, \bibinfo{year}{2019}.
\newblock \bibinfo{title}{Inverting magnetotelluric responses in a three-dimensional earth using fast forward approximations based on artificial neural networks}.
\newblock \bibinfo{journal}{Computers \& Geosciences} \bibinfo{volume}{127}, \bibinfo{pages}{44--52}.
\newblock \URLprefix \url{https://www.sciencedirect.com/science/article/pii/S009830041830462X}, \DOIprefix\doi{10.1016/j.cageo.2019.03.002}.
\bibitem[{Cranmer(2023)}]{cranmer_interpretable_2023}
\bibinfo{author}{Cranmer, M.}, \bibinfo{year}{2023}.
\newblock \bibinfo{title}{Interpretable {Machine} {Learning} for {Science} with {PySR} and {SymbolicRegression}.jl}.
\newblock \URLprefix \url{http://arxiv.org/abs/2305.01582}. \bibinfo{note}{arXiv:2305.01582 [astro-ph, physics:physics]}.
\bibitem[{Egbert et~al.(2022)Egbert, Yang, Bedrosian, Key, Livelybrooks, Schultz, Kelbert and Parris}]{egbert_fluid_2022}
\bibinfo{author}{Egbert, G.D.}, \bibinfo{author}{Yang, B.}, \bibinfo{author}{Bedrosian, P.A.}, \bibinfo{author}{Key, K.}, \bibinfo{author}{Livelybrooks, D.W.}, \bibinfo{author}{Schultz, A.}, \bibinfo{author}{Kelbert, A.}, \bibinfo{author}{Parris, B.}, \bibinfo{year}{2022}.
\newblock \bibinfo{title}{Fluid transport and storage in the {Cascadia} forearc influenced by overriding plate lithology}.
\newblock \bibinfo{journal}{Nature Geoscience} \bibinfo{volume}{15}, \bibinfo{pages}{677--682}.
\newblock \URLprefix \url{https://www.nature.com/articles/s41561-022-00981-8}, \DOIprefix\doi{10.1038/s41561-022-00981-8}.
\bibitem[{Gao et~al.(2018)Gao, Zhang, Zhang, Chen, Cheng, Jia, Li, Fu, Gao and Xin}]{gao_three-dimensional_2018}
\bibinfo{author}{Gao, J.}, \bibinfo{author}{Zhang, H.}, \bibinfo{author}{Zhang, S.}, \bibinfo{author}{Chen, X.}, \bibinfo{author}{Cheng, Z.}, \bibinfo{author}{Jia, X.}, \bibinfo{author}{Li, S.}, \bibinfo{author}{Fu, L.}, \bibinfo{author}{Gao, L.}, \bibinfo{author}{Xin, H.}, \bibinfo{year}{2018}.
\newblock \bibinfo{title}{Three-dimensional magnetotelluric imaging of the geothermal system beneath the {Gonghe} {Basin}, {Northeast} {Tibetan} {Plateau}}.
\newblock \bibinfo{journal}{Geothermics} \bibinfo{volume}{76}, \bibinfo{pages}{15--25}.
\newblock \URLprefix \url{https://www.sciencedirect.com/science/article/pii/S0375650518300245}, \DOIprefix\doi{10.1016/j.geothermics.2018.06.009}.
\bibitem[{Guo et~al.(2020)Guo, Egbert, Dong and Wei}]{guo_modular_2020}
\bibinfo{author}{Guo, Z.}, \bibinfo{author}{Egbert, G.}, \bibinfo{author}{Dong, H.}, \bibinfo{author}{Wei, W.}, \bibinfo{year}{2020}.
\newblock \bibinfo{title}{Modular finite volume approach for {3D} magnetotelluric modeling of the {Earth} medium with general anisotropy}.
\newblock \bibinfo{journal}{Physics of the Earth and Planetary Interiors} \bibinfo{volume}{309}, \bibinfo{pages}{106585}.
\newblock \URLprefix \url{https://linkinghub.elsevier.com/retrieve/pii/S0031920120302260}, \DOIprefix\doi{10.1016/j.pepi.2020.106585}.
\bibitem[{Jiang et~al.(2022)Jiang, Duan, Doublier, Clark, Schofield, Brodie and Goodwin}]{jiang_application_2022}
\bibinfo{author}{Jiang, W.}, \bibinfo{author}{Duan, J.}, \bibinfo{author}{Doublier, M.}, \bibinfo{author}{Clark, A.}, \bibinfo{author}{Schofield, A.}, \bibinfo{author}{Brodie, R.C.}, \bibinfo{author}{Goodwin, J.}, \bibinfo{year}{2022}.
\newblock \bibinfo{title}{Application of multiscale magnetotelluric data to mineral exploration: an example from the east {Tennant} region, {Northern} {Australia}}.
\newblock \bibinfo{journal}{Geophysical Journal International} \bibinfo{volume}{229}, \bibinfo{pages}{1628--1645}.
\newblock \URLprefix \url{https://doi.org/10.1093/gji/ggac029}, \DOIprefix\doi{10.1093/gji/ggac029}.
\bibitem[{Kolmogorov(1957)}]{kolmogorov_representation_1957}
\bibinfo{author}{Kolmogorov, A.K.}, \bibinfo{year}{1957}.
\newblock \bibinfo{title}{On the {Representation} of {Continuous} {Functions} of {Several} {Variables} by {Superposition} of {Continuous} {Functions} of {One} {Variable} and {Addition}}.
\newblock \bibinfo{journal}{Doklady Akademii Nauk SSSR} \bibinfo{volume}{114}, \bibinfo{pages}{369--373}.
\bibitem[{Kontolati et~al.(2024)Kontolati, Goswami, Em~Karniadakis and Shields}]{kontolati_learning_2024}
\bibinfo{author}{Kontolati, K.}, \bibinfo{author}{Goswami, S.}, \bibinfo{author}{Em~Karniadakis, G.}, \bibinfo{author}{Shields, M.D.}, \bibinfo{year}{2024}.
\newblock \bibinfo{title}{Learning nonlinear operators in latent spaces for real-time predictions of complex dynamics in physical systems}.
\newblock \bibinfo{journal}{Nature Communications} \bibinfo{volume}{15}, \bibinfo{pages}{5101}.
\newblock \URLprefix \url{https://www.nature.com/articles/s41467-024-49411-w}, \DOIprefix\doi{10.1038/s41467-024-49411-w}.
\bibitem[{Kovachki et~al.(2024)Kovachki, Li, Liu, Azizzadenesheli, Bhattacharya, Stuart and Anandkumar}]{kovachki_neural_2024}
\bibinfo{author}{Kovachki, N.}, \bibinfo{author}{Li, Z.}, \bibinfo{author}{Liu, B.}, \bibinfo{author}{Azizzadenesheli, K.}, \bibinfo{author}{Bhattacharya, K.}, \bibinfo{author}{Stuart, A.}, \bibinfo{author}{Anandkumar, A.}, \bibinfo{year}{2024}.
\newblock \bibinfo{title}{Neural {Operator}: {Learning} {Maps} {Between} {Function} {Spaces}}.
\newblock \URLprefix \url{http://arxiv.org/abs/2108.08481}, \DOIprefix\doi{10.5555/3648699.3648788}. \bibinfo{note}{arXiv:2108.08481 [cs, math]}.
\bibitem[{Li(2024)}]{li_kolmogorov-arnold_2024}
\bibinfo{author}{Li, Z.}, \bibinfo{year}{2024}.
\newblock \bibinfo{title}{Kolmogorov-{Arnold} {Networks} are {Radial} {Basis} {Function} {Networks}}.
\newblock \URLprefix \url{http://arxiv.org/abs/2405.06721}. \bibinfo{note}{arXiv:2405.06721 [cs]}.
\bibitem[{Li et~al.(2020)Li, Kovachki, Azizzadenesheli, Liu, Bhattacharya, Stuart and Anandkumar}]{li_neural_2020}
\bibinfo{author}{Li, Z.}, \bibinfo{author}{Kovachki, N.}, \bibinfo{author}{Azizzadenesheli, K.}, \bibinfo{author}{Liu, B.}, \bibinfo{author}{Bhattacharya, K.}, \bibinfo{author}{Stuart, A.}, \bibinfo{author}{Anandkumar, A.}, \bibinfo{year}{2020}.
\newblock \bibinfo{title}{Neural {Operator}: {Graph} {Kernel} {Network} for {Partial} {Differential} {Equations}}.
\newblock \URLprefix \url{http://arxiv.org/abs/2003.03485}. \bibinfo{note}{arXiv:2003.03485 [cs, math, stat]}.
\bibitem[{Li et~al.(2021)Li, Kovachki, Azizzadenesheli, Liu, Bhattacharya, Stuart and Anandkumar}]{li_fourier_2021}
\bibinfo{author}{Li, Z.}, \bibinfo{author}{Kovachki, N.}, \bibinfo{author}{Azizzadenesheli, K.}, \bibinfo{author}{Liu, B.}, \bibinfo{author}{Bhattacharya, K.}, \bibinfo{author}{Stuart, A.}, \bibinfo{author}{Anandkumar, A.}, \bibinfo{year}{2021}.
\newblock \bibinfo{title}{Fourier {Neural} {Operator} for {Parametric} {Partial} {Differential} {Equations}}.
\newblock \URLprefix \url{http://arxiv.org/abs/2010.08895}. \bibinfo{note}{arXiv:2010.08895 [cs, math]}.
\bibitem[{Liu et~al.(2021)Liu, Lin, Cao, Hu, Wei, Zhang, Lin and Guo}]{liu_swin_2021}
\bibinfo{author}{Liu, Z.}, \bibinfo{author}{Lin, Y.}, \bibinfo{author}{Cao, Y.}, \bibinfo{author}{Hu, H.}, \bibinfo{author}{Wei, Y.}, \bibinfo{author}{Zhang, Z.}, \bibinfo{author}{Lin, S.}, \bibinfo{author}{Guo, B.}, \bibinfo{year}{2021}.
\newblock \bibinfo{title}{Swin {Transformer}: {Hierarchical} {Vision} {Transformer} using {Shifted} {Windows}}, in: \bibinfo{booktitle}{2021 {IEEE}/{CVF} {International} {Conference} on {Computer} {Vision} ({ICCV})}, \bibinfo{publisher}{IEEE}, \bibinfo{address}{Montreal, QC, Canada}. pp. \bibinfo{pages}{9992--10002}.
\newblock \URLprefix \url{https://ieeexplore.ieee.org/document/9710580/}, \DOIprefix\doi{10.1109/ICCV48922.2021.00986}.
\bibitem[{Liu et~al.(2024)Liu, Wang, Vaidya, Ruehle, Halverson, Soljačić, Hou and Tegmark}]{liu_kan_2024}
\bibinfo{author}{Liu, Z.}, \bibinfo{author}{Wang, Y.}, \bibinfo{author}{Vaidya, S.}, \bibinfo{author}{Ruehle, F.}, \bibinfo{author}{Halverson, J.}, \bibinfo{author}{Soljačić, M.}, \bibinfo{author}{Hou, T.Y.}, \bibinfo{author}{Tegmark, M.}, \bibinfo{year}{2024}.
\newblock \bibinfo{title}{{KAN}: {Kolmogorov}-{Arnold} {Networks}}.
\newblock \URLprefix \url{https://arxiv.org/abs/2404.19756v4}.
\bibitem[{Lu et~al.(2021)Lu, Jin, Pang, Zhang and Karniadakis}]{lu_learning_2021}
\bibinfo{author}{Lu, L.}, \bibinfo{author}{Jin, P.}, \bibinfo{author}{Pang, G.}, \bibinfo{author}{Zhang, Z.}, \bibinfo{author}{Karniadakis, G.E.}, \bibinfo{year}{2021}.
\newblock \bibinfo{title}{Learning nonlinear operators via {DeepONet} based on the universal approximation theorem of operators}.
\newblock \bibinfo{journal}{Nature Machine Intelligence} \bibinfo{volume}{3}, \bibinfo{pages}{218--229}.
\newblock \URLprefix \url{https://www.nature.com/articles/s42256-021-00302-5}, \DOIprefix\doi{10.1038/s42256-021-00302-5}. \bibinfo{note}{publisher: Nature Publishing Group}.
\bibitem[{Newman and Alumbaugh(2000)}]{newman_three-dimensional_2000}
\bibinfo{author}{Newman, G.A.}, \bibinfo{author}{Alumbaugh, D.L.}, \bibinfo{year}{2000}.
\newblock \bibinfo{title}{Three-dimensional magnetotelluric inversion using non-linear conjugate gradients}.
\newblock \bibinfo{journal}{Geophysical Journal International} \bibinfo{volume}{140}, \bibinfo{pages}{410--424}.
\newblock \URLprefix \url{https://doi.org/10.1046/j.1365-246x.2000.00007.x}, \DOIprefix\doi{10.1046/j.1365-246x.2000.00007.x}.
\bibitem[{Peng et~al.(2023)Peng, Yang, Liu and Xu}]{peng_rapid_2023}
\bibinfo{author}{Peng, Z.}, \bibinfo{author}{Yang, B.}, \bibinfo{author}{Liu, L.}, \bibinfo{author}{Xu, Y.}, \bibinfo{year}{2023}.
\newblock \bibinfo{title}{Rapid surrogate modeling of magnetotelluric in the frequency domain using physics-driven deep neural networks}.
\newblock \bibinfo{journal}{Computers \& Geosciences} \bibinfo{volume}{176}, \bibinfo{pages}{105360}.
\newblock \URLprefix \url{https://linkinghub.elsevier.com/retrieve/pii/S009830042300064X}, \DOIprefix\doi{10.1016/j.cageo.2023.105360}.
\bibitem[{Peng et~al.(2022)Peng, Yang, Xu, Wang, Liu and Zhang}]{peng_rapid_2022}
\bibinfo{author}{Peng, Z.}, \bibinfo{author}{Yang, B.}, \bibinfo{author}{Xu, Y.}, \bibinfo{author}{Wang, F.}, \bibinfo{author}{Liu, L.}, \bibinfo{author}{Zhang, Y.}, \bibinfo{year}{2022}.
\newblock \bibinfo{title}{Rapid {Surrogate} {Modeling} of {Electromagnetic} {Data} in {Frequency} {Domain} {Using} {Neural} {Operator}}.
\newblock \bibinfo{journal}{IEEE Transactions on Geoscience and Remote Sensing} \bibinfo{volume}{60}, \bibinfo{pages}{1--12}.
\newblock \DOIprefix\doi{10.1109/TGRS.2022.3222507}. \bibinfo{note}{conference Name: IEEE Transactions on Geoscience and Remote Sensing}.
\bibitem[{Raissi et~al.(2019)Raissi, Perdikaris and Karniadakis}]{raissi_physics-informed_2019}
\bibinfo{author}{Raissi, M.}, \bibinfo{author}{Perdikaris, P.}, \bibinfo{author}{Karniadakis, G.E.}, \bibinfo{year}{2019}.
\newblock \bibinfo{title}{Physics-informed neural networks: {A} deep learning framework for solving forward and inverse problems involving nonlinear partial differential equations}.
\newblock \bibinfo{journal}{Journal of Computational Physics} \bibinfo{volume}{378}, \bibinfo{pages}{686--707}.
\newblock \URLprefix \url{https://www.sciencedirect.com/science/article/pii/S0021999118307125}, \DOIprefix\doi{10.1016/j.jcp.2018.10.045}.
\bibitem[{Romano et~al.(2014)Romano, Balasco, Lapenna, Siniscalchi, Telesca and Tripaldi}]{romano_sensitivity_2014}
\bibinfo{author}{Romano, G.}, \bibinfo{author}{Balasco, M.}, \bibinfo{author}{Lapenna, V.}, \bibinfo{author}{Siniscalchi, A.}, \bibinfo{author}{Telesca, L.}, \bibinfo{author}{Tripaldi, S.}, \bibinfo{year}{2014}.
\newblock \bibinfo{title}{On the sensitivity of long-term magnetotelluric monitoring in {Southern} {Italy} and source-dependent robust single station transfer function variability}.
\newblock \bibinfo{journal}{Geophysical Journal International} \bibinfo{volume}{197}, \bibinfo{pages}{1425--1441}.
\newblock \URLprefix \url{https://doi.org/10.1093/gji/ggu083}, \DOIprefix\doi{10.1093/gji/ggu083}.
\bibitem[{Shan et~al.(2022)Shan, Guo, Li, Yang, Xu and Liang}]{shan_application_2022}
\bibinfo{author}{Shan, T.}, \bibinfo{author}{Guo, R.}, \bibinfo{author}{Li, M.}, \bibinfo{author}{Yang, F.}, \bibinfo{author}{Xu, S.}, \bibinfo{author}{Liang, L.}, \bibinfo{year}{2022}.
\newblock \bibinfo{title}{Application of {Multitask} {Learning} for 2-{D} {Modeling} of {Magnetotelluric} {Surveys}: {TE} {Case}}.
\newblock \bibinfo{journal}{IEEE Transactions on Geoscience and Remote Sensing} \bibinfo{volume}{60}, \bibinfo{pages}{1--9}.
\newblock \URLprefix \url{https://ieeexplore.ieee.org/document/9513472/}, \DOIprefix\doi{10.1109/TGRS.2021.3101119}.
\bibitem[{Shukla et~al.(2024)Shukla, Toscano, Wang, Zou and Karniadakis}]{shukla_comprehensive_2024}
\bibinfo{author}{Shukla, K.}, \bibinfo{author}{Toscano, J.D.}, \bibinfo{author}{Wang, Z.}, \bibinfo{author}{Zou, Z.}, \bibinfo{author}{Karniadakis, G.E.}, \bibinfo{year}{2024}.
\newblock \bibinfo{title}{A comprehensive and {FAIR} comparison between {MLP} and {KAN} representations for differential equations and operator networks}.
\newblock \URLprefix \url{http://arxiv.org/abs/2406.02917}. \bibinfo{note}{arXiv:2406.02917 [physics]}.
\bibitem[{Siripunvaraporn and Egbert(2009)}]{siripunvaraporn_wsinv3dmt_2009}
\bibinfo{author}{Siripunvaraporn, W.}, \bibinfo{author}{Egbert, G.}, \bibinfo{year}{2009}.
\newblock \bibinfo{title}{{WSINV3DMT}: {Vertical} magnetic field transfer function inversion and parallel implementation}.
\newblock \bibinfo{journal}{Physics of the Earth and Planetary Interiors} \bibinfo{volume}{173}, \bibinfo{pages}{317--329}.
\newblock \URLprefix \url{https://www.sciencedirect.com/science/article/pii/S0031920109000144}, \DOIprefix\doi{10.1016/j.pepi.2009.01.013}.
\bibitem[{Smith(2014)}]{smith_electromagnetic_2014}
\bibinfo{author}{Smith, R.}, \bibinfo{year}{2014}.
\newblock \bibinfo{title}{Electromagnetic {Induction} {Methods} in {Mining} {Geophysics} from 2008 to 2012}.
\newblock \bibinfo{journal}{Surveys in Geophysics} \bibinfo{volume}{35}, \bibinfo{pages}{123--156}.
\newblock \URLprefix \url{http://link.springer.com/10.1007/s10712-013-9227-1}, \DOIprefix\doi{10.1007/s10712-013-9227-1}.
\bibitem[{SS et~al.(2024)SS, AR, R and KP}]{ss_chebyshev_2024}
\bibinfo{author}{SS, S.}, \bibinfo{author}{AR, K.}, \bibinfo{author}{R, G.}, \bibinfo{author}{KP, A.}, \bibinfo{year}{2024}.
\newblock \bibinfo{title}{Chebyshev {Polynomial}-{Based} {Kolmogorov}-{Arnold} {Networks}: {An} {Efficient} {Architecture} for {Nonlinear} {Function} {Approximation}}.
\newblock \URLprefix \url{http://arxiv.org/abs/2405.07200}. \bibinfo{note}{arXiv:2405.07200 [cs]}.
\bibitem[{{Tianping Chen} and {Hong Chen}(1995)}]{tianping_chen_universal_1995}
\bibinfo{author}{{Tianping Chen}}, \bibinfo{author}{{Hong Chen}}, \bibinfo{year}{1995}.
\newblock \bibinfo{title}{Universal approximation to nonlinear operators by neural networks with arbitrary activation functions and its application to dynamical systems}.
\newblock \bibinfo{journal}{IEEE Transactions on Neural Networks} \bibinfo{volume}{6}, \bibinfo{pages}{911--917}.
\newblock \URLprefix \url{http://ieeexplore.ieee.org/document/392253/}, \DOIprefix\doi{10.1109/72.392253}.
\bibitem[{Varilsuha and Candansayar(2018)}]{varilsuha_3d_2018}
\bibinfo{author}{Varilsuha, D.}, \bibinfo{author}{Candansayar, M.E.}, \bibinfo{year}{2018}.
\newblock \bibinfo{title}{{3D} magnetotelluric modeling by using finite-difference method: {Comparison} study of different forward modeling approaches}.
\newblock \bibinfo{journal}{GEOPHYSICS} \bibinfo{volume}{83}, \bibinfo{pages}{WB51--WB60}.
\newblock \URLprefix \url{https://library.seg.org/doi/abs/10.1190/geo2017-0406.1}, \DOIprefix\doi{10.1190/geo2017-0406.1}. \bibinfo{note}{publisher: Society of Exploration Geophysicists}.
\bibitem[{Wang et~al.(2024a)Wang, Jiang, Deng, Wang, Yang and Yuan}]{wang_three_2024}
\bibinfo{author}{Wang, X.}, \bibinfo{author}{Jiang, P.}, \bibinfo{author}{Deng, F.}, \bibinfo{author}{Wang, S.}, \bibinfo{author}{Yang, R.}, \bibinfo{author}{Yuan, C.}, \bibinfo{year}{2024}a.
\newblock \bibinfo{title}{Three {Dimensional} {Magnetotelluric} {Forward} {Modeling} {Through} {Deep} {Learning}}.
\newblock \bibinfo{journal}{IEEE Transactions on Geoscience and Remote Sensing} , \bibinfo{pages}{1--1}\URLprefix \url{https://ieeexplore.ieee.org/abstract/document/10530923}, \DOIprefix\doi{10.1109/TGRS.2024.3401587}. \bibinfo{note}{conference Name: IEEE Transactions on Geoscience and Remote Sensing}.
\bibitem[{Wang et~al.(2024b)Wang, Guo, Liu, Li, Liu, Chen, Cao, Yin and Cao}]{wang_divergence-free_2024}
\bibinfo{author}{Wang, Y.}, \bibinfo{author}{Guo, R.}, \bibinfo{author}{Liu, J.}, \bibinfo{author}{Li, J.}, \bibinfo{author}{Liu, R.}, \bibinfo{author}{Chen, H.}, \bibinfo{author}{Cao, X.}, \bibinfo{author}{Yin, Z.}, \bibinfo{author}{Cao, C.}, \bibinfo{year}{2024}b.
\newblock \bibinfo{title}{A divergence-free vector finite-element method for efficient {3D} magnetotelluric forward modeling}.
\newblock \bibinfo{journal}{GEOPHYSICS} \bibinfo{volume}{89}, \bibinfo{pages}{E1--E11}.
\newblock \URLprefix \url{https://library.seg.org/doi/10.1190/geo2023-0037.1}, \DOIprefix\doi{10.1190/geo2023-0037.1}.
\bibitem[{Yu et~al.(2020)Yu, Unsworth, Wang, Li, Wang, Li, Hu and Cai}]{yu_new_2020}
\bibinfo{author}{Yu, N.}, \bibinfo{author}{Unsworth, M.}, \bibinfo{author}{Wang, X.}, \bibinfo{author}{Li, D.}, \bibinfo{author}{Wang, E.}, \bibinfo{author}{Li, R.}, \bibinfo{author}{Hu, Y.}, \bibinfo{author}{Cai, X.}, \bibinfo{year}{2020}.
\newblock \bibinfo{title}{New {Insights} {Into} {Crustal} and {Mantle} {Flow} {Beneath} the {Red} {River} {Fault} {Zone} and {Adjacent} {Areas} on the {Southern} {Margin} of the {Tibetan} {Plateau} {Revealed} by a 3‐{D} {Magnetotelluric} {Study}}.
\newblock \bibinfo{journal}{Journal of Geophysical Research: Solid Earth} \bibinfo{volume}{125}, \bibinfo{pages}{e2020JB019396}.
\newblock \URLprefix \url{https://agupubs.onlinelibrary.wiley.com/doi/10.1029/2020JB019396}, \DOIprefix\doi{10.1029/2020JB019396}.
\bibitem[{Yu et~al.(2025)Yu, Wang, Chen and Kong}]{yu_two-dimensional_2025}
\bibinfo{author}{Yu, N.}, \bibinfo{author}{Wang, C.}, \bibinfo{author}{Chen, H.}, \bibinfo{author}{Kong, W.}, \bibinfo{year}{2025}.
\newblock \bibinfo{title}{A two-dimensional magnetotelluric deep learning inversion approach based on improved {Dense} {Convolutional} {Network}}.
\newblock \bibinfo{journal}{Computers \& Geosciences} \bibinfo{volume}{194}, \bibinfo{pages}{105765}.
\newblock \URLprefix \url{https://www.sciencedirect.com/science/article/pii/S0098300424002486}, \DOIprefix\doi{10.1016/j.cageo.2024.105765}.
\bibitem[{Yu et~al.(2024a)Yu, Zhang, Chen, Kong, Feng and Qian}]{yu_three-dimensional_2024}
\bibinfo{author}{Yu, N.}, \bibinfo{author}{Zhang, H.}, \bibinfo{author}{Chen, H.}, \bibinfo{author}{Kong, W.}, \bibinfo{author}{Feng, X.}, \bibinfo{author}{Qian, Y.}, \bibinfo{year}{2024}a.
\newblock \bibinfo{title}{Three-{Dimensional} {Unstructured} {Finite} {Element} {Modeling} of {Magnetotelluric} {Problems} {Allowing} for {Continuous} {Variation} of {Conductivity} in {Each} {Block}}.
\newblock \bibinfo{journal}{IEEE Transactions on Geoscience and Remote Sensing} \bibinfo{volume}{62}, \bibinfo{pages}{1--12}.
\newblock \URLprefix \url{https://ieeexplore.ieee.org/document/10526292/}, \DOIprefix\doi{10.1109/TGRS.2024.3398601}.
\bibitem[{Yu et~al.(2024b)Yu, Yu and Wang}]{yu_kan_2024}
\bibinfo{author}{Yu, R.}, \bibinfo{author}{Yu, W.}, \bibinfo{author}{Wang, X.}, \bibinfo{year}{2024}b.
\newblock \bibinfo{title}{{KAN} or {MLP}: {A} {Fairer} {Comparison}}.
\newblock \URLprefix \url{http://arxiv.org/abs/2407.16674}. \bibinfo{note}{arXiv:2407.16674 [cs]}.
\bibitem[{Zhang et~al.(2020)Zhang, Hu, Jin, Deng, Wu and Chen}]{zhang_maxwells_2020}
\bibinfo{author}{Zhang, P.}, \bibinfo{author}{Hu, Y.}, \bibinfo{author}{Jin, Y.}, \bibinfo{author}{Deng, S.}, \bibinfo{author}{Wu, X.}, \bibinfo{author}{Chen, J.}, \bibinfo{year}{2020}.
\newblock \bibinfo{title}{A {Maxwell}'s {Equations} {Based} {Deep} {Learning} {Method} for {Time} {Domain} {Electromagnetic} {Simulations}}, in: \bibinfo{booktitle}{2020 {IEEE} {Texas} {Symposium} on {Wireless} and {Microwave} {Circuits} and {Systems} ({WMCS})}, pp. \bibinfo{pages}{1--4}.
\newblock \URLprefix \url{https://ieeexplore.ieee.org/document/9172407}, \DOIprefix\doi{10.1109/WMCS49442.2020.9172407}.
\bibitem[{Zhu et~al.(2022)Zhu, Liu, Cui and Gong}]{zhu_scalable_2022}
\bibinfo{author}{Zhu, X.}, \bibinfo{author}{Liu, J.}, \bibinfo{author}{Cui, Y.}, \bibinfo{author}{Gong, C.}, \bibinfo{year}{2022}.
\newblock \bibinfo{title}{A {Scalable} {Parallel} {Algorithm} for 3-{D} {Magnetotelluric} {Finite} {Element} {Modeling} in {Anisotropic} {Media}}.
\newblock \bibinfo{journal}{IEEE Transactions on Geoscience and Remote Sensing} \bibinfo{volume}{60}, \bibinfo{pages}{1--14}.
\newblock \URLprefix \url{https://ieeexplore.ieee.org/document/9438942/}, \DOIprefix\doi{10.1109/TGRS.2021.3078735}.

\end{thebibliography}

\end{document}